\documentclass[aps,prb,twocolumn]{revtex4-2}
\usepackage{amsmath,amssymb}
\usepackage{graphicx}
\usepackage{dcolumn}
\usepackage{mathtools}
\usepackage{bm}
\usepackage{subcaption}
\usepackage{float}
\usepackage{color}

\usepackage{indentfirst}
\usepackage{indentfirst}
\usepackage{xcolor}
\usepackage{array}
\usepackage{makecell}
\usepackage{tabularx}

\begin{document}
	\title {Variation of Transition Energy of an Off-Center $D^0$	Impurity in a Gaussian Quantum Dot with Spin-Orbit Interactions and External Fields}

	
	\author{Rajdip Banerjee$^{1}$}
	
	\author{Pooja Saini$^{1,2}$}
	\thanks{Corresponding author.\\
		E-mail address: jnu.poojasaini@gmail.com}
	\author{Arnab Bhowmik$^{1}$}	
	\author{Dwipesh Majumder$^{1}$}
	
	
	\affiliation{$^{1}$Department of Physics, Indian Institute of Engineering Science and Technology, Howrah, West Bengal-711103, India}
	
	\affiliation{$^{2}$Department of Mathematics, Indian Institute of Engineering Science and Technology, Howrah, West Bengal-711103, India}

	\begin{abstract}
 Simultaneous effects of electric and magnetic fields on an off-centered $D^0$
 impurity in a GaAs quantum dot with Gaussian confinement are investigated using the variational approach, including Rashba and Dresselhaus spin-orbit interactions. The ground (1s) and first excited (2p) states are analysed to determine their energy spectra and impurity binding energies. Fields and spin-orbit effects on both energies are examined for varying impurity positions. The transition energy decreases concavely with electric field, while radial impurity displacement further lowers it. The magnetic response is analysed through magnetization and susceptibility, revealing diamagnetic behaviour in both states. Its dependence on spin-orbit interaction strengths, impurity position, and electric-field strength is further explored. For the parameter ranges considered, the electric and magnetic field strengths span upto $10\mathrm{kV/cm}$ and $ 10\mathrm{T}$ respectively, while the calculated transition energy and state-dependent binding energies lie within approximately $[10, 25]\,R_y^*$ and $[1.4, 4.5]\,R_y^*$, respectively. These results show that impurity off-centering enables tuning the excited-state properties of Gaussian quantum dots using external fields and spin-orbit interactions.
		
	\end{abstract}
	\maketitle
	\date{\today}
	\section{Introduction}
Donor impurities in semiconductor quantum dots provide a useful framework for examining localized carrier states under quantum confinement. In particular, hydrogenic \(D^0\) centeres offer a convenient model for studying how impurity position, confinement, and external fields modify the electronic structure of low-dimensional semiconductor systems.
In an early study, Brus developed a theoretical framework for size-dependent excited states, optical transitions, and redox properties of semiconductor crystallites \cite{Brus1984}. Subsequent studies established the role of quantum confinement in controlling exciton correlations, energies, and oscillator strengths, along with enhanced biexciton stability in ultrasmall quantum dots \cite{Bryant1988,Bryant1990}. Optical anticrossings due to coupling between centre-of-mass and relative exciton motion were also reported in parabolic dots \cite{Halonen1992}, while effective-mass calculations for InAs/GaAs dots showed good agreement with experimental results \cite{Marzin1994}. Further investigations highlighted the influence of confinement and material parameters on electron correlations and pair localization in Gaussian quantum dots \cite{Boyacioglu2007,Boyacioglu2013}. In donor-impurity systems, enhanced binding energies in smaller dots and the competing effects of Rashba and Dresselhaus interactions on donor binding were demonstrated \cite{SAINI2020106641, Mukhopadhyay2023}.
  \\
  Beyond the electronic and magnetic properties, impurity-related effects also extend to the nonlinear optical response of quantum dots. Hasanirokh and Naifar investigated harmonic generation in impurity-containing cylindrical core/shell/shell quantum dots (CSSQDs), demonstrating the influence of magnetic fields, dielectric environment, structural dimensions, Rashba coupling, and impurity position on nonlinear optical responses \cite{Hasanirokh2023}. In a related study, Kria \textit{et al.} examined Rashba induced modifications of the linear and nonlinear optical properties of cylindrical CSSQDs, showing that magnetic fields shift optical resonances towards higher energies, while stronger Rashba coupling enhances absorption coefficients and refractive-index response \cite{Kria2022}. 
 \\
  \indent Spin-orbit interactions, arising from the coupling between an electron's orbital motion and spin, play a crucial role in semiconductor nanostructures by lifting spin degeneracy and enabling electrical control of spin states. In GaAs-based semiconductor nanostructures, spin-orbit coupling is predominantly governed by two mechanisms: the Rashba and Dresselhaus interactions. The Rashba interaction arises primarily from structural inversion asymmetry and can be effectively controlled by an externally applied electric field, whereas the Dresselhaus interaction is an intrinsic consequence of the bulk inversion asymmetry of the crystal structure \cite{Rashba1960,Dresselhaus1955}.
  In quantum dots, Rashba and Dresselhaus spin-orbit interactions significantly influence the electronic structure, impurity binding energies, and spin-dependent transitions \cite{Manchon2015, RastegarSedehi2026}. Voskoboynikov et al. examined the magnetic behaviour of parabolic quantum dots with few-electrons in the presence of Rashba spin-orbit interaction(RSOI). Their study revealed that spin-orbit coupling significantly alters the magnetization and magnetic susceptibility through spin-split energy-level crossings at low magnetic fields\cite{Voskoboynikov2003}.
 \\ \indent Extending spin-orbit investigations to impurity-containing systems, Elsaid \textit{et al.} examined the influence of Rashba coupling and external electric and magnetic fields on the magnetic and thermal properties of donor-containing parabolic GaAs quantum dots \cite{Elsaid2020}. 
  This study highlights the importance of examining impurity-induced modifications in semiconductor quantum dots. Among these impurities, hydrogenic donors have received considerable theoretical attention, particularly regarding their binding characteristics in spherical quantum dots.
 Porras-Montenegro and Perez-Merchancano reported distinct radius-dependent donor binding in finite and infinite spherical GaAs-(Ga,Al)As quantum dots, including a critical radius maximum for finite wells \cite{PorrasMontenegro1992}, while Deng \textit{et al.} showed that spatially varying dielectric screening strengthens impurity binding, particularly for acceptors in smaller dots \cite{Deng1994}. Bella and Navaneethakrishnan reported size- and orbital-dependent stability of excited \(2p\) donor states, with negligible spin-orbit corrections \cite{Bella2004}. Sahoo \textit{et al.} identified an anomalous Stark response and found that magnetic fields shift resonances to higher energies while narrowing them \cite{Sahoo2008}. Meanwhile, Rejo Jeice \textit{et al.} showed that confined-state energies rise with temperature but decrease with increasing dot size and conduction-band non-parabolicity \cite{Jeice2015}.  
 In particular, on-center donor impurities have received considerable attention because the high spatial symmetry of the confinement considerably simplifies the theoretical formulation while capturing the essential features of impurity-induced electronic states \cite{Yang1998}.
\\ \indent Considerable effort has been devoted to understanding the response of donor impurities in semiconductor quantum dots to externally applied magnetic and electric fields.
 Along similar lines, Ben Hammou \textit{et al.} demonstrated the dependence of donor diamagnetic susceptibility in pyramidal CSSQDs on confinement geometry and electric-field strength, particularly under weak confinement \cite{BenHammou2026}, and further examined field-dependent donor binding by incorporating polaronic effects and conduction-band non-parabolicity \cite{BenHammou2026b}. Earlier, Jahan \textit{et al.} explored excitonic properties in two- and three-dimensional GaAs Gaussian quantum dots under magnetic fields, demonstrating the sensitivity of binding energy, spatial extent, and oscillator strength to confinement and field strength; their shifted \(1/N\)-expansion results also agreed closely with exact diagonalization \cite{Jahan2018}. 
In this context, Dane \textit{et al.} employed an effective-mass variational approach to investigate an on-centred \(D^0\) donor in an infinitely confined spherical GaAs/AlAs quantum dot, demonstrating that binding energy decreases with electric field but increases with magnetic field, with a pronounced variation near a critical dot radius \cite{Dane2008}. Subsequent studies by Akbas \textit{et al.} and Dane \textit{et al.} separately examined the electric- and magnetic-field effects, establishing their respective influences on donor binding \cite{Akbas2009,Dane2010}. The combined effects of magnetic fields and spin-orbit interactions in Gaussian-confined GaAs quantum dots also have been investigated, revealing stronger suppression of donor binding by Rashba coupling than by Dresselhaus coupling, whereas magnetic fields enhance binding and modify the magnetic properties \cite{Saini2019}.
In another recent study, Arraoui \textit{et al.} showed that donor binding and polarizability in GaAs four-quantum-dot systems depend on impurity position and confinement geometry, with polarizability varying non-monotonically under weak confinement \cite{Arraoui2025}, while studies of double quantum dots revealed electric-field and impurity-position effects on binding and self-polarization, alongside magnetic field induced changes in binding energy and photoionization \cite{Arraoui2022,Arraoui2024}. Investigations of triple quantum wires revealed additional control through impurity location and inter-wire coupling, with implications for binding energy and polarizability \cite{Arraoui2026}. In double parabolic quantum dots, Kasapoglu \textit{et al.} demonstrated that impurity configuration and electric fields can modify electronic-level hybridization and induce anticrossing features \cite{Kasapoglu2026}.
\\ \indent Experimental investigations have provided important insight into the tunability of semiconductor quantum dots through external fields. Using Stark spectroscopy, Warburton \textit{et al.} established a connection between the electric-field response of the dipole moment and polarizability and the geometrical characteristics of self-assembled quantum dots \cite{Warburton2002}. In a separate investigation, Tang \textit{et al.} demonstrated through photoluminescence measurements that the excitonic charge configuration of individual InAs/GaAs quantum dots can be regulated by simultaneously applied electric and magnetic fields \cite{Tang2014}.
\\ \indent On the other hand, Donor displacement provides an additional means of controlling confined electronic states by breaking the spatial symmetry of the quantum dot. Kachu \textit{et al.} demonstrated that its interplay with spin-orbit coupling modifies both ground and excited-state properties in Gaussian quantum dots \cite{Kachu2024}. Similarly, Saini \textit{et al.} reported substantial changes in ground-state energy and magnetic response arising from impurity displacement and confinement asymmetry \cite{Saini2026}.
\\ \indent Despite these developments, the combined response of an off-centred donor under simultaneous electric and magnetic fields, with Rashba and Dresselhaus interactions treated together, remains insufficiently explored, particularly for the first excited \(2p\) state and its transition and magnetic properties. Motivated by this gap, we examine how impurity displacement, external electric and magnetic fields, and Rashba-Dresselhaus coupling jointly influence the \(1s\) and \(2p\) donor states, with particular emphasis on their energies, binding characteristics, \(1s\rightarrow2p\) transition, and magnetic response.
\\
The remainder of this manuscript is organized as follows. Section II presents the theoretical formulation and variational methodology. Section III discusses the numerical results, focusing on the effects of impurity displacement, external fields, and spin-orbit interactions on the electronic and magnetic properties. Finally, Section IV summarizes the principal findings and conclusions. 

\vspace{0.5cm}
\section{MODEL AND METHOD}
We consider a hydrogenic \(D^0\) donor embedded away from the symmetry centre of a three-dimensional Gaussian quantum dot, with the confined electron exposed simultaneously to electric and magnetic fields. Rashba and Dresselhaus spin-orbit contributions are included to examine their combined influence on the impurity states. The corresponding Hamiltonian within the effective-mass approximation, widely used in several studies of spherical\cite{Dane2008,Dane2010} as well as Gaussian\cite{Saini2019,Saini2021,Saini2026,BANERJEE2026174510} quantum dots, is given by,
    \begin{equation}\label{eq:ham_1}
    	H=[\frac{1}{2m^*}(\mathbf{p}+\frac{e}{c}\mathbf{A})^2 -\frac{e^2}{\epsilon \lvert \mathbf{r}-\mathbf{D} \rvert}+V(\mathbf{r})]I+H_{R}+H_{D}+H_{F}
    \end{equation}
    where,
    \vspace{-0.2cm}
    \begin{equation}\label{eq:rsoi}
    	H_{R}=\frac{\alpha_{R}}{\hbar}[\boldsymbol{\sigma} \times (\mathbf{p}+\frac{e}{c}\mathbf{A})]_{z}
    \end{equation}	
    \vspace{-0.2cm}
    \begin{equation}\label{eq:dsoi}
    	H_{D}=\frac{\beta_{D}}{\hbar}[\sigma_{x}(p_{x}+\frac{e}{c}A_{x})-\sigma_{y}(p_{y}+\frac{e}{c}A_{y})]
    \end{equation}
	\begin{equation}\label{eq:efield}
		H_{F}=e\;\mathbf{F}\cdot\mathbf{r}
	\end{equation}
	
In Eq.-\ref{eq:ham_1}, the first term describes a hydrogenic donor ($D^0$) impurity confined by a Gaussian potential in the presence of an external magnetic field(B)\cite{Mukhopadhyay2023,Saini2019,Kachu2024},\cite{Nowak2011}. Here, $m^*$ and $e$ denote the effective mass and electron's charge, respectively, while $\mathbf{r}=(x,y,z)$ and $p$ represent the electron position and canonical momentum operator. The magnetic field is incorporated through the vector potential $A$, satisfying $\mathbf{B}=\boldsymbol{\nabla} \times \mathbf{A}$, and $\epsilon$ is the dielectric constant of the host material. The second term, describes the off-centered hydrogenic donor impurity where $D$ is the measure of radial displacement of the impurity from the quantum dot center. The confinement potential is characterized as $V(r)=-V_{0}e^{-\frac{r^2}{2R^2}}$  with the $R$ describing the effective quantum-dot radius. The symbol I denotes an identity matrix of $2\times2$ dimension. The second and third terms correspond to the Rashba\cite{Rashba1960} and Dresselhaus\cite{Dresselhaus1955} spin-orbit interactions, where $\sigma_{x}$,$\sigma_{y}$ and $\sigma_{z}$ are the Pauli matrices, and $\alpha_{R}$ and $\beta_{D}$ represent their respective coupling constants. The final term accounts for the applied electric field along $z$ direction\cite{BenHammou2026,BenHammou2026b,Arraoui2025}.	Throughout our calculations, we have considered the symmetric gauge choice\cite{Bhuiyan2020}, $\mathbf{A}=\frac{B}{2}(-y,x,0)$. The spin-orbit interaction is incorporated by applying the unitary operator, $U=e^{iS}$ with $
S=\frac{m^{*}}{\hbar^{2}}
\left[
\alpha_{R}\left(y\sigma_{x}-x\sigma_{y}\right)
+\beta_{D}\left(x\sigma_{x}-y\sigma_{y}\right)
\right].
$ This approach, first proposed by Aleiner and Falco \cite{Aleiner2001} and later employed in related studies \cite{ValinRodriguez2002,Kumar2016}, facilitates the diagonalization of the Hamiltonian. The transformed Hamiltonian, subsequently expanding in powers of $\alpha_{R}$and $\beta_{D}$, with terms retained up to second order, is given by,
\begin{equation}
	\begin{aligned}
		\tilde{\mathcal{H}}
		&=
		\left(
		\frac{p^{2}}{2m^{*}}
		+\frac{1}{8}m^{*}\omega_{c}^{2}\rho^{2}
		-\frac{e^{2}}{\varepsilon \lvert r-D \rvert}
		-V_{0}e^{-\frac{r^{2}}{2R^{2}}}
		\right)I
		\\[2mm]
		&\quad
		-\frac{m^{*}}{\hbar^{2}}
		\left(\alpha_{R}^{2}+\beta_{D}^{2}\right)I
		-\frac{m^{*}}{\hbar^{3}}
		\left(\alpha_{R}^{2}-\beta_{D}^{2}\right)\sigma_{z}L_{z}
		+\frac{\omega_{c}}{2}L_{z}
		\\[2mm]
		&\quad
		-\frac{m^{*}}{2\hbar^{3}}
		\left(\alpha_{R}^{2}-\beta_{D}^{2}\right)
		\omega_{c}\sigma_{z}\rho^{2}
		+eFr\cos\theta .
	\end{aligned}
\end{equation} 
where, $\omega_{c}=(eB/m^{*}c)$ , $\rho^2=(x^2+y^2)$ and $L_{z}=-i\hbar(\frac{\partial}{\partial \phi}).$ The electronic states are obtained by minimizing the variational energy functional with respect to the parameters entering the trial functions. The forms adopted for the two impurity states are:
\begin{equation}
	\psi_{1s}(r)=N_{1}e^{-\alpha r^{2}-\beta (r-D)-\gamma r\cos\theta-im\phi}
\end{equation}
\vspace{-0.6cm}
\begin{equation}
	\psi_{2p}(r)=N_{2}\lambda\sin\theta(r\cos\theta-D) e^{-\delta r^2-\mu (r-D)-im\phi}
\end{equation}
Where, $\psi_{1s}$ is the trial wavefunction for the ground state$(m=0)$ and $\psi_{2p}$ is that of the first excites state with $m=1$. Here, $N_{1}$,$\;$$N_{2}$ are normalization constants for ground and first excited state wavefunction respectively. $\alpha,\;\beta$ and $\gamma$ are the variational parameters for ground state wavefunction whereas $\lambda$,$\delta$ and $\mu$ are variational parameters for 1st excited state wavefunction.
In the ground-state wavefunction, $\alpha$ and $\beta$ govern the radial localization, whereas $\gamma$ allows an asymmetric spatial distribution along the electric-field direction. For the first excited state, $\delta$ and $\mu$ regulate the radial extent and donor-related localization, while the angular factor$(\lambda)$ determines the nodal structure. The magnetic-field dependence enters through the vector potential in the Hamiltonian, including the associated orbital Zeeman and diamagnetic contributions. Consequently, the optimized variational parameters depend on the applied field strengths, allowing the electronic states to adjust their spatial distributions in response to external perturbations. Related variational descriptions of off-centred donor states in Gaussian quantum dots have been employed in earlier theoretical investigations \cite{Kachu2024,BANERJEE2026174510}
Following the variational prescription adopted in earlier theoretical treatments \cite{Dane2008,Dane2010}, we determine the ground-state energy of the donor complex through the optimization of the normalized Hamiltonian expectation value,
\begin{equation}
	E_{1s}(D^0) = \min_{\alpha,\beta,\gamma}
	\left[
	\frac{\left\langle \psi_{1s}(\mathbf{r}) \middle|\tilde{\mathcal{H}} \middle| \psi_{1s}(\mathbf{r}) \right\rangle}
	{\left\langle \psi_{1s}(\mathbf{r}) \middle| \psi_{1s}(\mathbf{r}) \right\rangle}
	\right]
\end{equation}
\begin{equation}
	E_{2p}(D^0) = \min_{\lambda,\delta,\mu}
	\left[
	\frac{\left\langle \psi_{2p}(\mathbf{r}) \middle|\tilde{\mathcal{H}} \middle| \psi_{2p}(\mathbf{r}) \right\rangle}
	{\left\langle \psi_{2p}(\mathbf{r}) \middle| \psi_{2p}(\mathbf{r}) \right\rangle}
	\right]
\end{equation}
For every combination of confinement characteristics and externally applied fields, the parameters $\alpha$, $\beta$, and $\gamma$ for ground state and $\lambda$,$\delta$ and $\mu$ for excited state are treated as independent optimization variables. The numerical search is initiated from several distinct starting configurations to avoid relying on a particular initial guess. Convergence is accepted when successive minimizations reproduce the same optimized energy within the numerical accuracy of Mathematica's NMinimize algorithm. The corresponding optimized parameter sets are subsequently employed to evaluate the electronic energy, donor binding energy, magnetic moment, and magnetic susceptibility of the ground$(1s)$ and the excited$(2p)$ state. The numerical procedures and computational settings adopted for the variational calculations are summarized in Table \ref{tab:mytable}.

\begin{table}[htbp]
	\centering
	\caption{Table of computational settings}
	\label{tab:mytable}
	
	\resizebox{\columnwidth}{!}{%
	\begin{tabular}{|c|c|}
		\hline
		\textbf{Computational aspect} & \textbf{Specification adopted} \\
		\hline
		Software & Wolfram Mathematica \\
		\hline
		Optimization routine & NMinimize \\
		\hline
		Ground-state variational parameters & $\alpha,\beta,\gamma$ \\
		\hline
		Excited-state variational parameters & $\lambda,\delta,\mu$ \\
		\hline
		Numerical integration routine & Integrate \\
		\hline
		Magnetic-field differentiation & Finite-difference. \\
		\hline
	\end{tabular}%
    }
\end{table}
  The transition energy of the $D^0$ donor complex, for the $2p\to1s$ transition of the off-centered donor impurity$(D^0)$ is then obtained from\cite{Boda2016},
\begin{equation}
	\Delta E(D^{0}) = E_{2p}(D^{0}) - E_{1s}(D^{0}),
\end{equation}
where $E_{1s}(D^0)$ is the energy of the $D^0$ impurity, corresponding to the ground$(1s)$ state and $E_{2p}(D^{0})$ is that corresponding to the first excited$(2p)$ state. 
\\
\indent The impurity binding energy for a given electronic state is defined\cite{Boda2016,BANERJEE2026174510} as the difference between the energy of the corresponding electron state in the absence of the impurity and that of the impurity-bound state, namely,
\begin{equation}
	\begin{aligned}
		E_{B}^{1s}(D^{0}) &= E_{1s}(e^{-})-E_{1s}(D^{0}),\\
		E_{B}^{2p}(D^{0}) &= E_{2p}(e^{-})-E_{2p}(D^{0}).
	\end{aligned}
\end{equation}
Here, $E_{1s}(e^{-})$ and $E_{2p}(e^{-})$ denote the energies of the impurity-free electron in the ground $(1s)$ and first excited $(2p)$ states, respectively. 
These impurity-free reference energies are calculated by excluding the Coulomb interaction between the donor and the electron, \(V_C(\mathbf r)=-e^2/(\epsilon|\mathbf r-\mathbf D|)\), from the Hamiltonian while determining the corresponding ground and first excited states. 
Whereas $E_{1s}(D^{0})$ and $E_{2p}(D^{0})$ represent the corresponding energies with the donor impurity confined in the same Gaussian quantum dot. The thermal stability of the impurity states is subsequently estimated using the condition $E_{B}=k_{B}T$, where $k_{B}$ is the Boltzmann constant and $T$ denotes the temperature. We define the relative binding energy as the difference between the binding energies of the first excited $(2p)$ and ground $(1s)$ states and it is expressed as,  
\begin{equation}
	\Delta E_{B} = E_{B}^{2p} - E_{B}^{1s}.
\end{equation}
Physically, this quantity measures the relative binding of the donor impurity between the two electronic states. A larger magnitude of $\Delta E_{B}$ indicates a stronger state-selective influence of the impurity on the electronic structure, whereas values approaching zero imply that the donor binds the ground and excited states nearly equally. 
\\
\indent The magnetic response of the quantum dot is subsequently analyzed by calculating the magnetic moment at zero-temperature defined\cite{Saini2019,BANERJEE2026174510} as, 
\begin{equation}
	M_{i}=-\frac{\partial E_{i}}{\partial B},
\end{equation}
and the corresponding magnetic susceptibility\cite{Saini2019,BANERJEE2026174510},
\begin{equation}
	\chi_{i}=\frac{\partial M_{i}}{\partial B}.
\end{equation}
where $ i=1s,2p$.
Both the $M_{i}$ and $\chi_{i}$  are determined for both ground and first excited state from the magnetic-field dependence of the total energy in that respective electronic state.
\\
Throughout this investigation, the electric-field strength is varied from \(0\) to \(10\,\mathrm{kV/cm}\), and the magnetic field is considered over the interval \(0\) to \(10\,\mathrm{T}\). These ranges are consistent with those adopted in several earlier theoretical investigations of semiconductor quantum dots subjected to external fields \cite{Zhang2011,Zeng2014,Saini2019,BANERJEE2026174510,Jahan2018,Dane2008,Dane2010}. An unrestricted increase in the applied electric field could induce excessive polarization of the donor-bound electron, potentially compromising its confinement within the quantum dot. The selected field interval therefore provides a physically meaningful regime for examining the interplay between electric-field-induced polarization, magnetic confinement, and spin-orbit coupling. Furthermore, the impurity displacement is restricted to \(D<R\), ensuring that the donor remains within the spatial extent of the confining structure throughout the investigation. 
\\
\indent The above formalism serves as the foundation for the numerical results presented in the following section, where the effects of external fields, impurity displacement, and spin-orbit interactions are systematically analyzed.
\subsection{Assumptions}
The present theoretical treatment relies on several assumptions that define the scope of the investigation. The system is considered to contain a single hydrogenic $(D^0)$ donor in a three-dimensional GaAs quantum dot with smooth Gaussian confinement. The electronic properties are described within the effective-mass approximation, using fixed material(GaAs) parameters for the effective mass and dielectric constant. A static and spatially uniform electric field is applied along the $z$-direction, and the magnetic field is incorporated along $z$-direction through the symmetric gauge. The Dresselhaus interaction, originating from bulk inversion asymmetry, holds two parts, one is linear in Dresselhaus parameter and the other one is cubic.
\[
H_D\propto
P_x(P_y^2-P_z^2)\sigma_x
+P_y(P_z^2-P_x^2)\sigma_y
+P_z(P_x^2-P_y^2)\sigma_z.
\]	
The cubic momentum contribution is negligiable for GaAs quantum dots. So, we have retained only the effective linear term. Accordingly,	
	\[
	H_D=\frac{\beta_D}{\hbar}
	\left[
	\sigma_x\left(p_x+\frac{e}{c}A_x\right)
	-\sigma_y\left(p_y+\frac{e}{c}A_y\right)
	\right],
	\]	
	where \(\beta_D=\gamma(\pi/a_0)^2\), with \(a_0\) denoting the quantum-dot height and \(\gamma=27.5\,\mathrm{eV\,\mathring A^3}\) representing the bulk Dresselhaus coefficient for GaAs. This assumption is widely used and Previous theoretical investigations have successfully employed this reduced formulation in describing the electronic properties of comparable quantum-dot structures\cite{Nowak2011,Saini2021,Mukhopadhyay2023,Kachu2024,BANERJEE2026174510}. To account for the spin-orbit contributions, we employ a unitary transformation of the Hamiltonian and retain terms up to second order in the Rashba and Dresselhaus coupling parameters, following an approach established in several earlier theoretical investigations \cite{Saini2019,Mukhopadhyay2023,Kachu2024,BANERJEE2026174510}. 

\section{NUMERICAL RESULTS}
The theoretical model developed above is now employed to investigate the electronic and magnetic properties of an off-centered $D^0$ impurity confined in a Gaussian quantum dot of GaAs. The numerical results presented below illustrate the combined effects of electric and magnetic fields, Rashba and Dresselhaus spin-orbit interactions, and impurity displacement on the ground- and excited-state energies, transition characteristics, binding energies, magnetic moment, and magnetic susceptibility. The suitable parameters chosen for GaAs is given by, $\epsilon=12.4$ and $m^{*}=0.067m_{0}$ \cite{Vurgaftman2001,Saini2019,Saini2021,Saini2026,BANERJEE2026174510}, where $m_{0}$ describes the bare mass of electron. These scale results $R^{*}_{y}=12 meV$ and $a_{B}^{*}=9.8  nm$. The SOI strengths($\alpha_{R}$ and $\beta_{D}$) are expressed in terms of $R_{y}^{*} a_{B}^{*}$ unless stated otherwise. The magnetic field strength$(B)$ is expressed in Tesla. The electric field strength$(F)$ is expressed in $kv/cm$.
\begin{figure}[htbp]
	\centering
	
     \includegraphics[width=\linewidth]{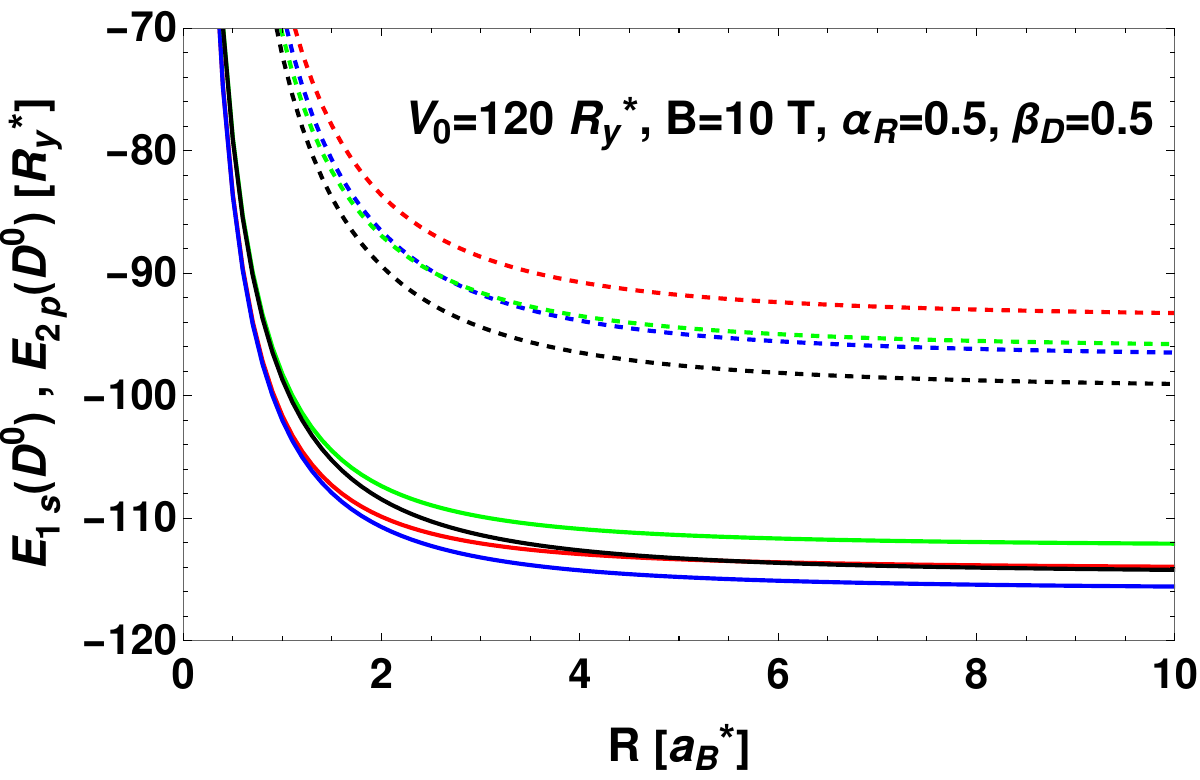}
	
	\caption{Ground($1s$) and First excited state($2p$) energy of the $D^0$ donor complex as a function of quantum dot radius($R$) for different values of impurity displacement$(D)$ and electric field strengths$(F)$. Continuous curves correspond to the \(1s\) state, whereas the dashed curves represent the \(2p\) state with $D=0.2,F=0.0$(Red),$D=0.2,F=6.0$(Blue),$D=0.8,F=0$(Green) and $D=0.8,F=6.0$(Black).}
	\label{fig:E_vs_R}
	
\end{figure}
\\
\indent Fig-\ref{fig:E_vs_R} describes the variation of both ground state$(1s)$ and first excited state$(2p)$ energy of the $D^0$ donor complex confined in a Gaussian quantum dot of GaAs, as a function of dot radius$(R)$, for different values of impurity position$(D)$ and electric field strength$(F)$. The solid lines denote the ground state$(GS)$ and the dashed lines denote the excited state$(ES)$ energy.  For both states, the energy decreases rapidly in the strong-confinement regime (small $R$) and gradually approaches a nearly constant value as the confinement weakens. This behaviour reflects the reduction of quantum confinement with increasing dot size. This type of behaviour was also reported in several earlier works\cite{Dane2008,Dane2010,Saini2019,BANERJEE2026174510}. It is also evident that an increase in the electric field shifts both energy levels to lower values, indicating enhanced localization of the electron. In contrast, the two states respond differently to impurity displacement. The \(1s\) level shifts upward as \(D\) increases, whereas the \(2p\) level moves in the opposite direction, with a stronger dependence on \(D\). The opposite trend of the $2p$ state can be traced by the distinct $D$-dependent structure of its variational wave function, containing the factor $(rcos\theta-D)$. As the impurity is displaced, this spatial structure allows the excited-state electron distribution to reorganize relative to the displaced donor and the confinement potential. The resulting changes in the Coulomb, confinement, and kinetic contributions lead to a net lowering of the $2p$ energy over the displacement range considered. The crossover of the curves of the ground state corresponding to $D=0.2,F=0.0$ and $D=0.8,F=6.0$ is due to the opposite effect of the electric field and impurity position on the energy. The modified behaviour of impurity position in excited state suppresses the crossover feature. 
\\ 
\begin{figure}[htbp]
	\centering
	
	\begin{minipage}[c]{0.47\linewidth}
		\centering
		\includegraphics[width=\linewidth,height=0.16\textheight]{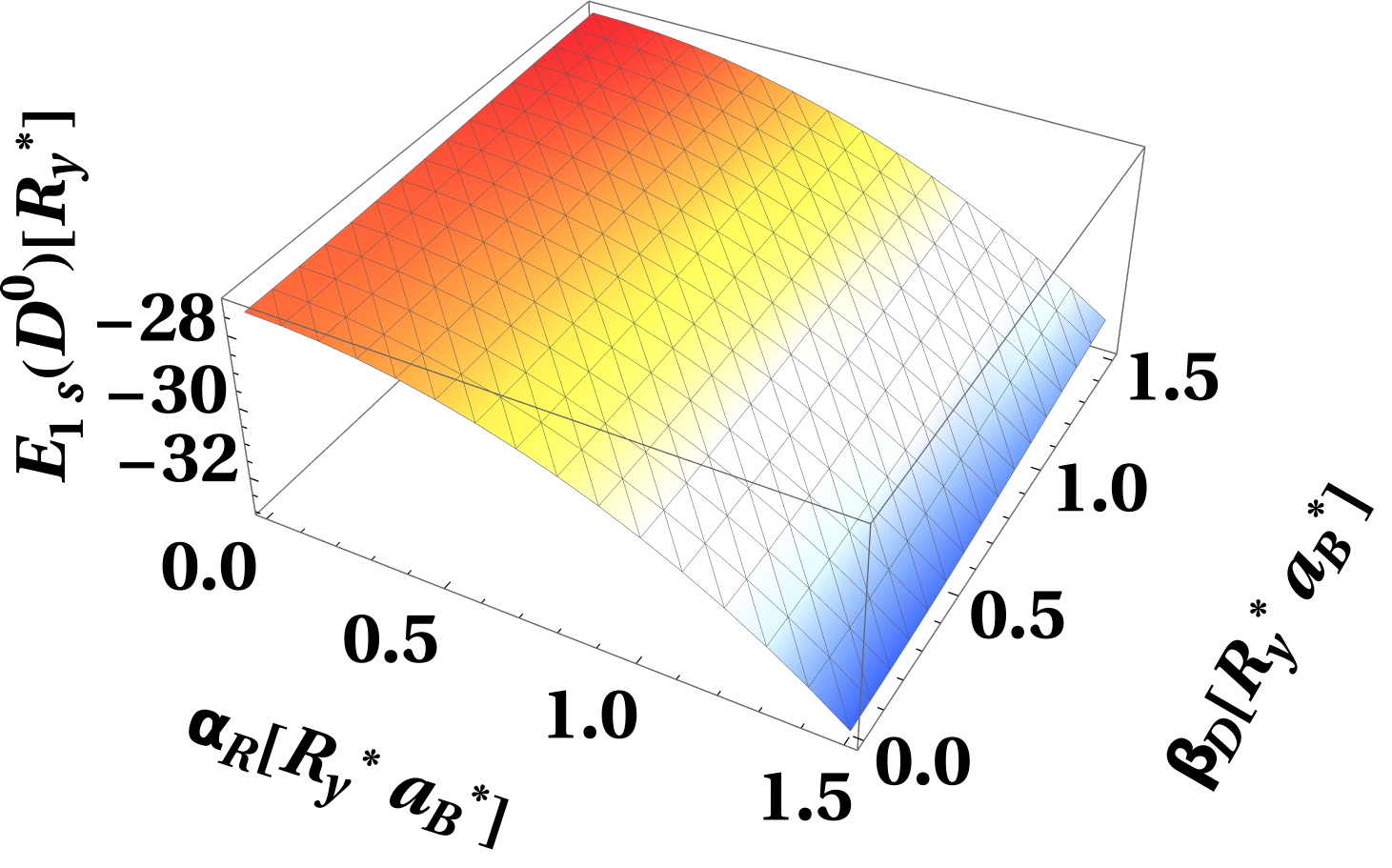}
		\vspace{2mm}
		{\small (a)}
	\end{minipage}
	\hfill
	\begin{minipage}[c]{0.47\linewidth}
		\centering
		\includegraphics[width=\linewidth,height=0.16\textheight]{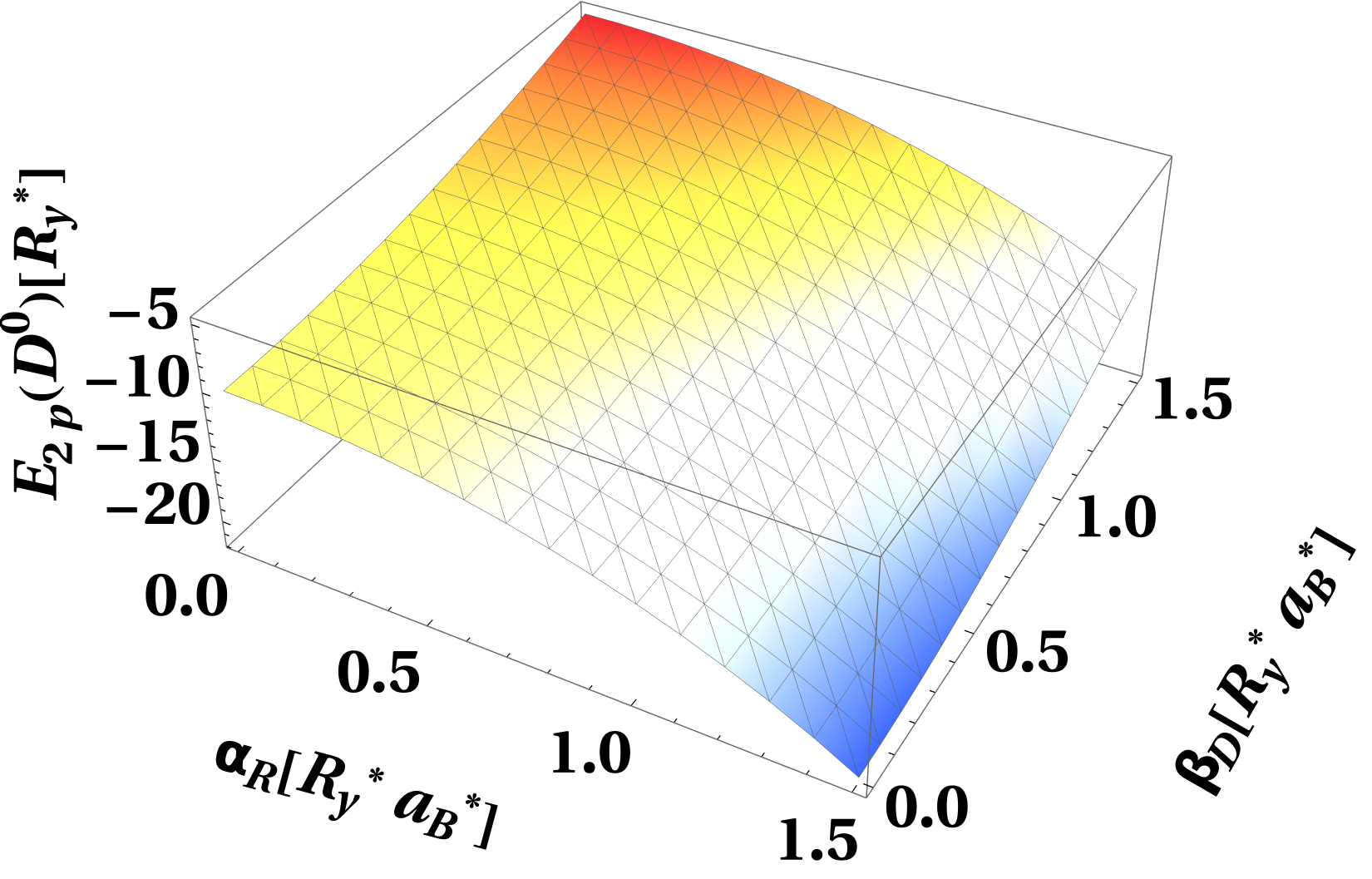}
		\vspace{2mm}
		{\small (b)}
	\end{minipage}

	\caption{3D plots illustrating the dependence of the (a) ground-state $(1s)$ and (b) first excited-state $(2p)$ energies of the off-centered $D^0$ impurity on the Rashba $(\alpha_{R})$ and Dresselhaus $(\beta_{D})$ spin-orbit interaction strengths with $V_{0}=40 R_{y}^{*}, B=8T, R=1.0 a_{B}^{*}, D=0.8 a_{B}^{*}, F=2.5 kv/cm$.}
	\label{fig:E_vs_ab}
	
\end{figure}

\indent Fig-\ref{fig:E_vs_ab} represents the variation of ground state(fig-\ref{fig:E_vs_ab}.a) and the first excited state(fig-\ref{fig:E_vs_ab}.b) energy of the off-centered hydrogenic $D^0$ donor complex$(D= 0.8)$  with RSOI$(\alpha_{R})$ and DSOI$(\beta_{D})$ strengths in the presence of simultaneous electric$(F=2.5 kv/cm)$ and magnetic field$(B=8 T)$. Both the ground state and first excited state energy decreases with RSOI strength$(\alpha_{R})$ but increases very slowly with DSOI strength$(\beta_{D})$. The increasing nature of excited state energy with $\beta_{D}$ is a little bit more prominent than that of the ground state. These results are in well compliance with the results reported earlier\cite{Kachu2024}. The different response to Dresselhaus coupling in the centered\cite{BANERJEE2026174510} and off-centered configurations indicates a strong dependence of the spin-orbit contribution on impurity position. In the centered case, the higher spatial symmetry of the system leads to a negative energy correction from both Rashba and Dresselhaus interactions. Once the impurity is displaced from the confinement center, this symmetry is reduced and the electron wave function undergoes a corresponding spatial redistribution. The resulting change in the expectation value of the momentum-dependent spin-orbit terms modifies their contributions to the total variational energy. For the off-centered configuration considered here, the Rashba contribution remains negative and comparatively dominant, whereas the Dresselhaus contribution becomes weakly positive. This behaviour is particularly noticeable for the $2p$ state because of its nodal and angular structure. The results therefore demonstrate that impurity off-centering can qualitatively modify the balance between the two spin-orbit interactions and, consequently, the energy spectrum of the donor complex.
\\
\begin{figure}[htbp]
	\centering
	
	\begin{minipage}[c]{0.48\linewidth}
		\centering
		\includegraphics[width=1.05\linewidth,height=0.16\textheight]{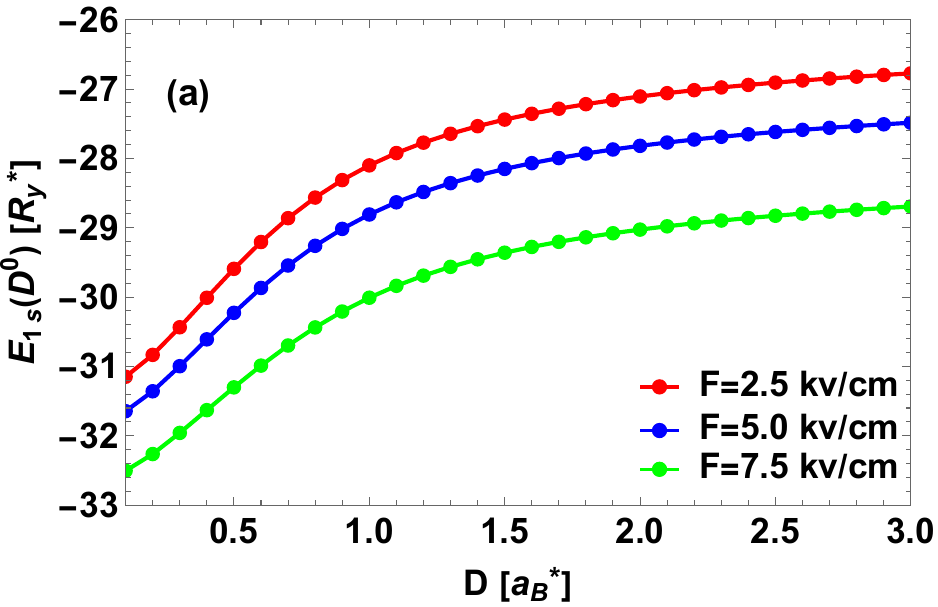}
	\end{minipage}
	\hfill
	\begin{minipage}[c]{0.48\linewidth}
		\centering
		\includegraphics[width=1.0\linewidth,height=0.16\textheight]{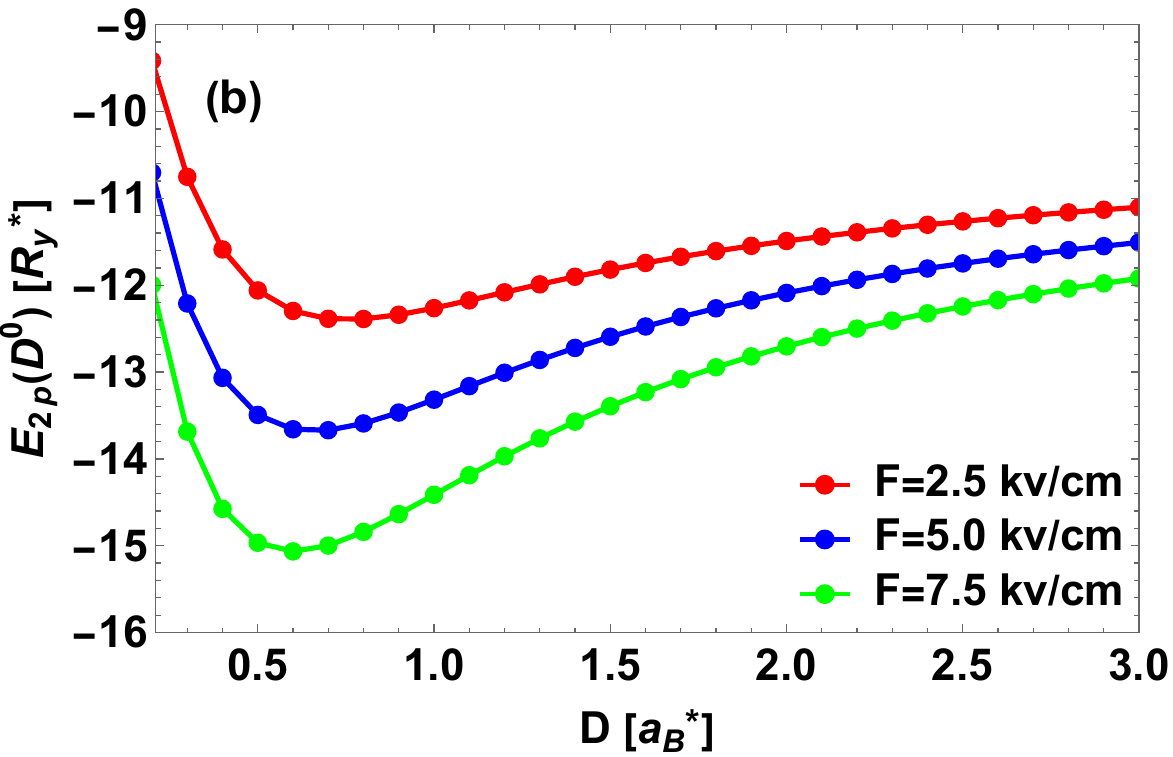}
	\end{minipage}
	
	\caption{Dependence of the (a) ground-state (1s) and (b) first excited-state (2p) energies of the off-centered $D^0$ impurity on the impurity displacement D for different electric field strengths$(F)$ with $V_{0}=40 R_{y}^{*}, B=7T, R=4 a_{B}^{*}, \alpha_{R}=0.5, \beta_{D}=0.5$. }
	\label{fig:E_vs_D}
	
\end{figure}
\indent Fig-\ref{fig:E_vs_D} illustrates the evolution of the ground $(1s)$(fig-\ref{fig:E_vs_D}.a) and first excited $(2p)$(fig-\ref{fig:E_vs_D}.b) state energies with impurity displacement for different electric field strengths. As the impurity moves away from the quantum-dot centere, the ground-state energy increases monotonically, indicating a gradual reduction in the electron-impurity Coulomb attraction due to the increasing separation between the electron and the donor. In contrast, the excited-state energy exhibits a non-monotonic behaviour, initially decreasing to a minimum before increasing at larger displacements. This distinct response originates from the more extended spatial distribution of the $2p$ wave function, whose overlap with the off-centered impurity varies more sensitively with displacement. Moreover, increasing the electric field shifts both energy levels towards lower values, with the effect being more pronounced for the excited state, reflecting its greater response to external perturbations. A similar reduction in the energy with increasing electric-field strength, even under the simultaneous application of a magnetic field, has also been reported for both spherical\cite{Dane2010} and Gaussian\cite{BANERJEE2026174510} quantum dots.
\begin{figure}[htbp]
	\centering
	\includegraphics[width=1.0\linewidth]{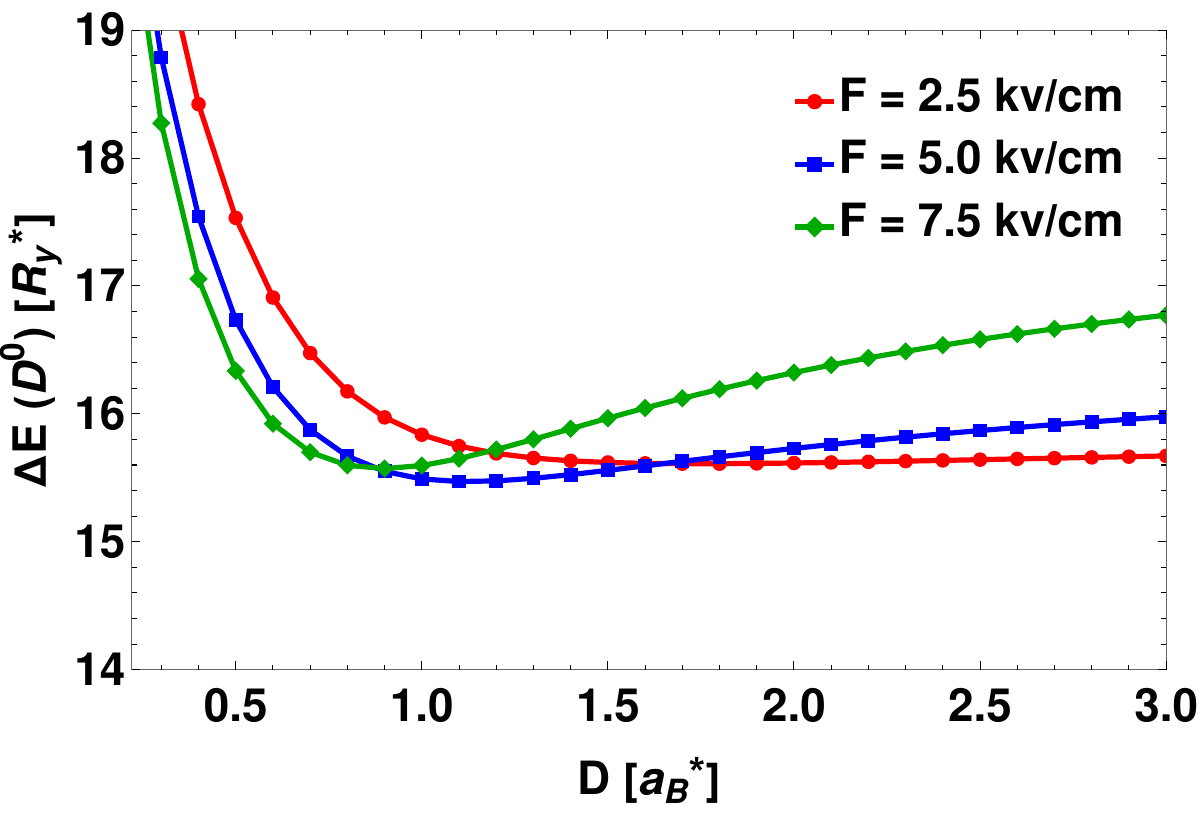}
	\caption{Dependence of the transition energy$(1s\to2p)$ of an off-centered $D^0$ impurity confined in a Gaussian quantum dot on the impurity displacement D for different values of the applied electric field strength$(F)$.}
	\label{fig:tran_en_vs_D}
\end{figure}
\\
\indent Fig-\ref{fig:tran_en_vs_D} presents the variation of the $1s\to2p$ transition energy with impurity displacement$(D)$ for different electric field$(F)$ strengths. For all values of the applied field, the transition energy decreases rapidly at small impurity displacements, reaches a minimum around $D=0.9a_{B}^{*}$, and then increases gradually as the impurity is moved further away from the dot center. The crossing behaviour arises from the different rates at which the ground and excited-state energies evolve with impurity displacement, resulting in a temporary reduction of the energy separation followed by its subsequent widening. Moreover, the increase in electric field produces a more pronounced enhancement of the transition energy at larger impurity displacements, indicating that the combined action of the electric field$(F)$ and impurity off-centering$(D)$ provides an effective means of tuning the inter-level transition.
\\
\begin{figure}[htbp]
	\centering
	
	\begin{minipage}[c]{0.47\linewidth}
		\centering
		\includegraphics[width=1.05\linewidth,height=0.14\textheight]{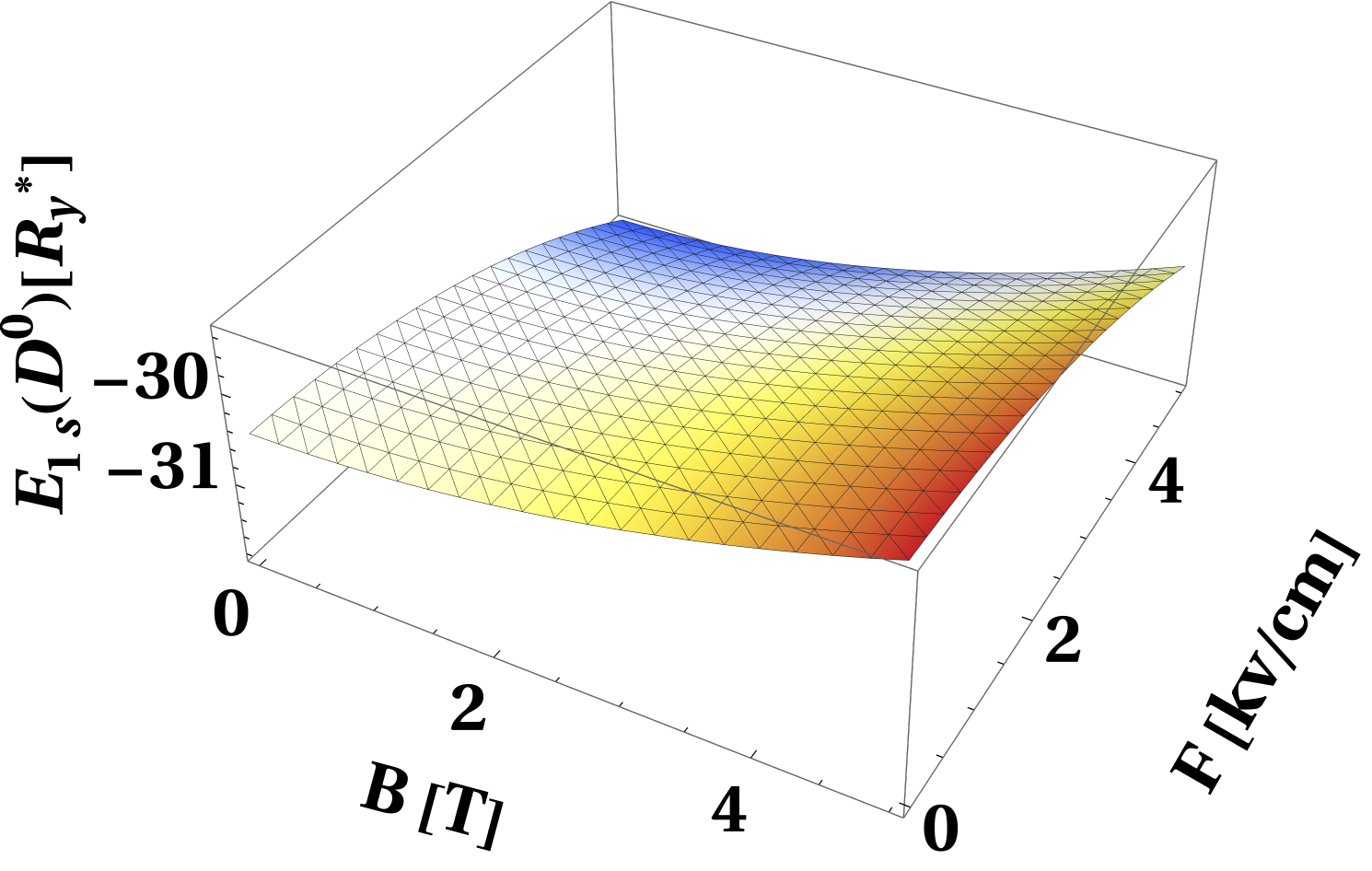}
		\vspace{2mm}
		{\small (a)}
	\end{minipage}
	\hfill
	\begin{minipage}[c]{0.48\linewidth}
		\centering
		\includegraphics[width=1.\linewidth,height=0.14\textheight]{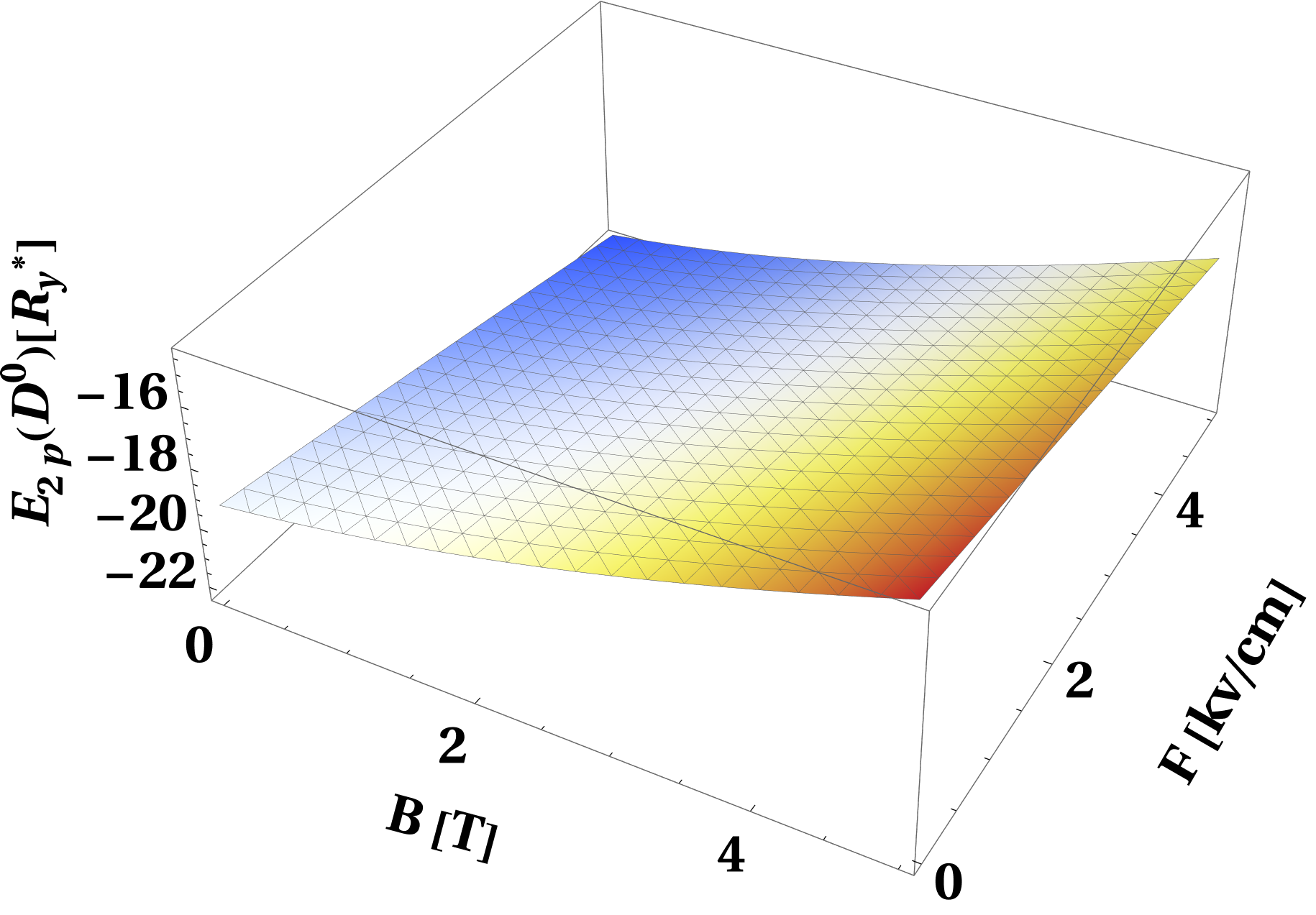}
		\vspace{2mm}
		{\small (b)}
	\end{minipage}
	\caption{Dependence of the ground-state energy \(E_{1s}\) (a) and first excited-state energy \(E_{2p}\) (b) on the applied magnetic field \(B\) and electric field \(F\) for an off-centred \(D^0\) donor in a Gaussian quantum dot with $V_{0}=40R_{y}^{*}, D=0.8a_{B}^{*}, R=1.0a_{B}^{*}, \alpha_{R}=0.5, \beta_{D}=0.5$. }
	\label{fig:E_vs_EB}
	
	\end{figure}
\indent Fig-\ref{fig:E_vs_EB} demonstrates the combined influence of the electric and magnetic fields on the ground $(1s)$ and first excited $(2p)$ state energies of the off-centered $D^0$ impurity. In both cases, the magnetic field$(B)$ lowers the energy, reflecting the enhanced localization of the electron under magnetic confinement. In contrast, the applied electric field$(F)$ shifts the energy towards higher values, with a considerably stronger effect on the excited state. Comparable magnetic-field-dependent trends in both ground- and excited-state energies have also been documented in earlier studies\cite{Jahan2018,Boda2016}, although those investigations considered the magnetic-field dependence independently, without incorporating a simultaneous electric field.
\\
\begin{figure}[htbp]
	\centering
	
	\begin{minipage}[c]{0.47\linewidth}
		\centering
		\includegraphics[width=1.05\linewidth,height=0.14\textheight]{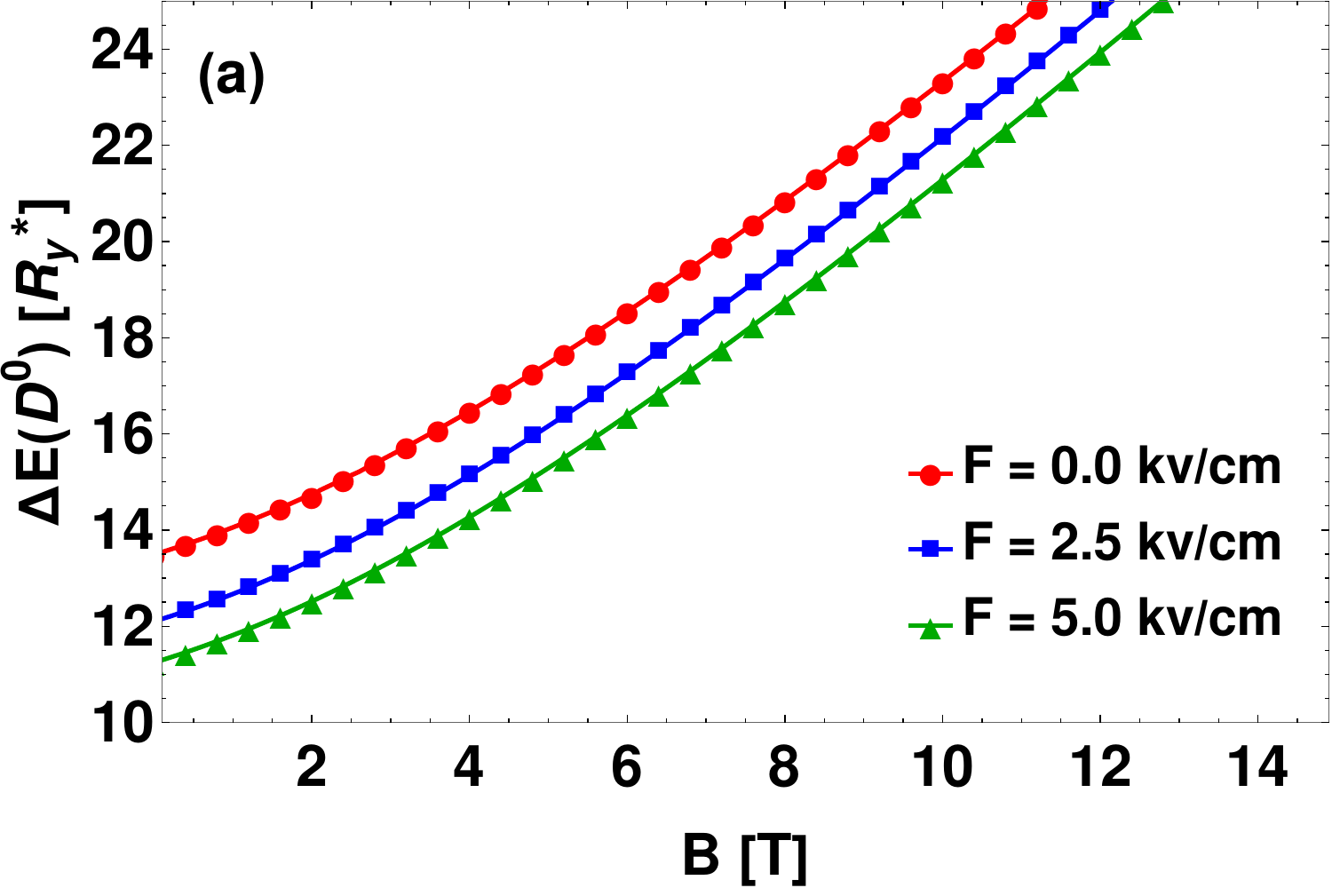}
		\vspace{2mm}
	\end{minipage}
	\hfill
	\begin{minipage}[c]{0.48\linewidth}
		\centering
		\includegraphics[width=1.\linewidth,height=0.14\textheight]{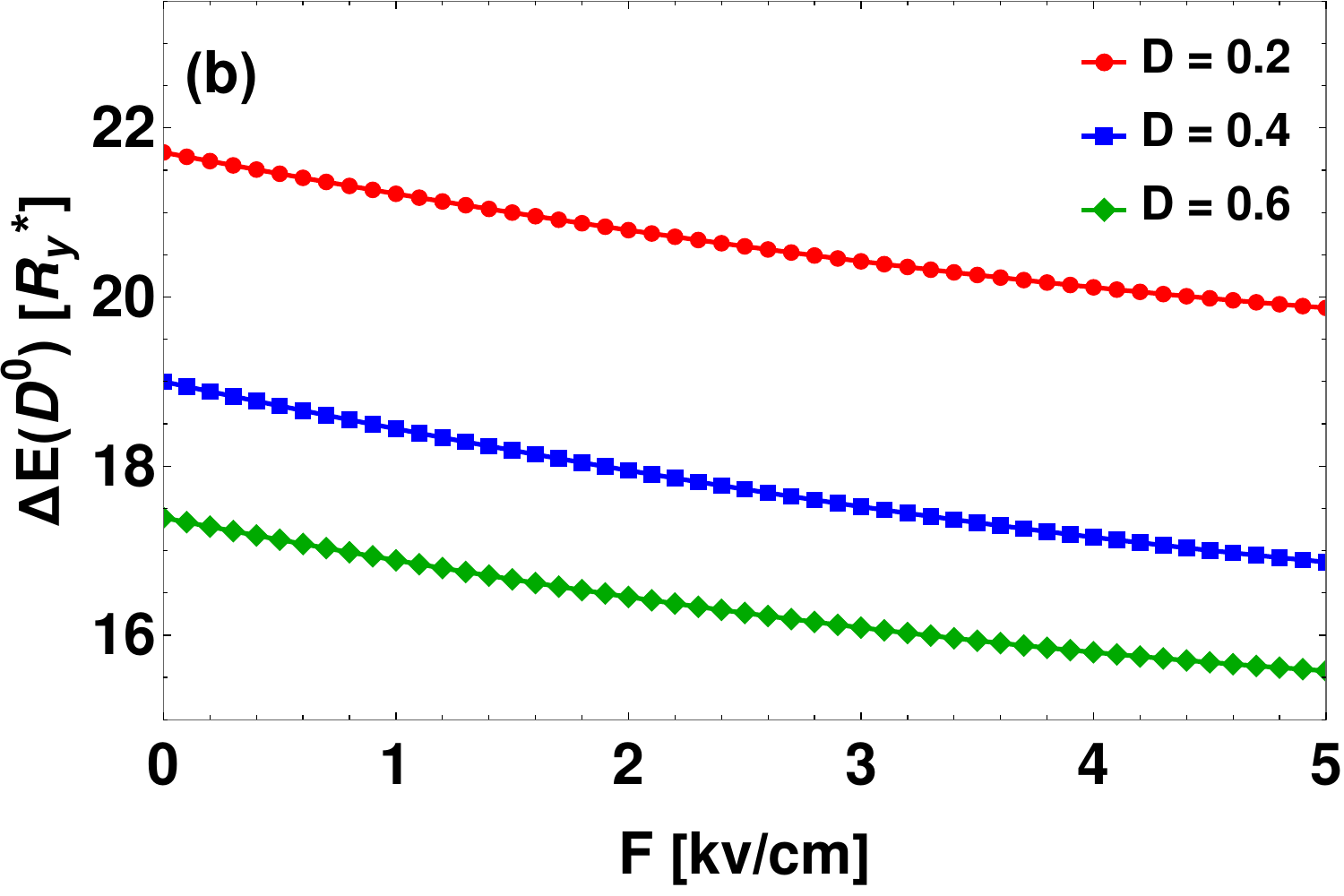}
		\vspace{2mm}
	\end{minipage}
	\caption{(a) Magnetic-field dependence of the 1s→2p transition energy of an off-centered $D^{0}$ impurity confined in a Gaussian quantum dot for different values of the applied electric field strength$(F)$ and (b) Variation of the $1s\to2p$ transition energy of the off-centered $D^0$ impurity with the applied electric field for different values of the impurity displacement with  $V_{0}=40R_{y}^{*}, R=1.0 a_{B}^{*}, D=0.4a_{B}^{*}, \alpha_{R}=0.5, \beta_{D}=0.5$.}
	\label{fig:tran_en_vs_EB}
	
\end{figure}
\\
\indent Fig-\ref{fig:tran_en_vs_EB}.(a) shows the variation of the 1s→2p transition energy with the applied magnetic field($B$) for different electric field strengths. The transition energy increases monotonically with increasing magnetic field, indicating that the energy separation between the ground and excited states becomes larger under stronger magnetic confinement. This behaviour originates from the magnetic-field-induced localization of the electronic wave functions, which affects the two states differently and consequently enlarges the transition energy. In contrast, increasing the electric field shifts the curves downward over the entire magnetic-field range, implying a reduction in the transition energy. The nearly parallel nature of the curves further suggests that the electric field primarily introduces an overall energy offset without significantly altering the magnetic-field dependence of the transition.
\\
\indent Fig-\ref{fig:tran_en_vs_EB}.(b) shows the dependence of the $1s\to2p$ transition energy$(\Delta E)$ on the applied electric field strength$(F)$ for different impurity displacements$(D )$. The transition energy decreases gradually with increasing electric field for a fixed value of impurity displacement$(D)$, indicating that the electric field reduces the energy separation between the ground and excited states. This reduction becomes more pronounced as the impurity is displaced farther from the dot center, leading to lower transition energies over the entire electric-field range. The nearly parallel nature of the curves suggests that the electric field influences the transition energy in a similar manner for different impurity locations, while the off-center displacement primarily determines the magnitude of the energy separation.
\begin{figure}[htbp]
	\centering
	\includegraphics[width=1.0\linewidth]{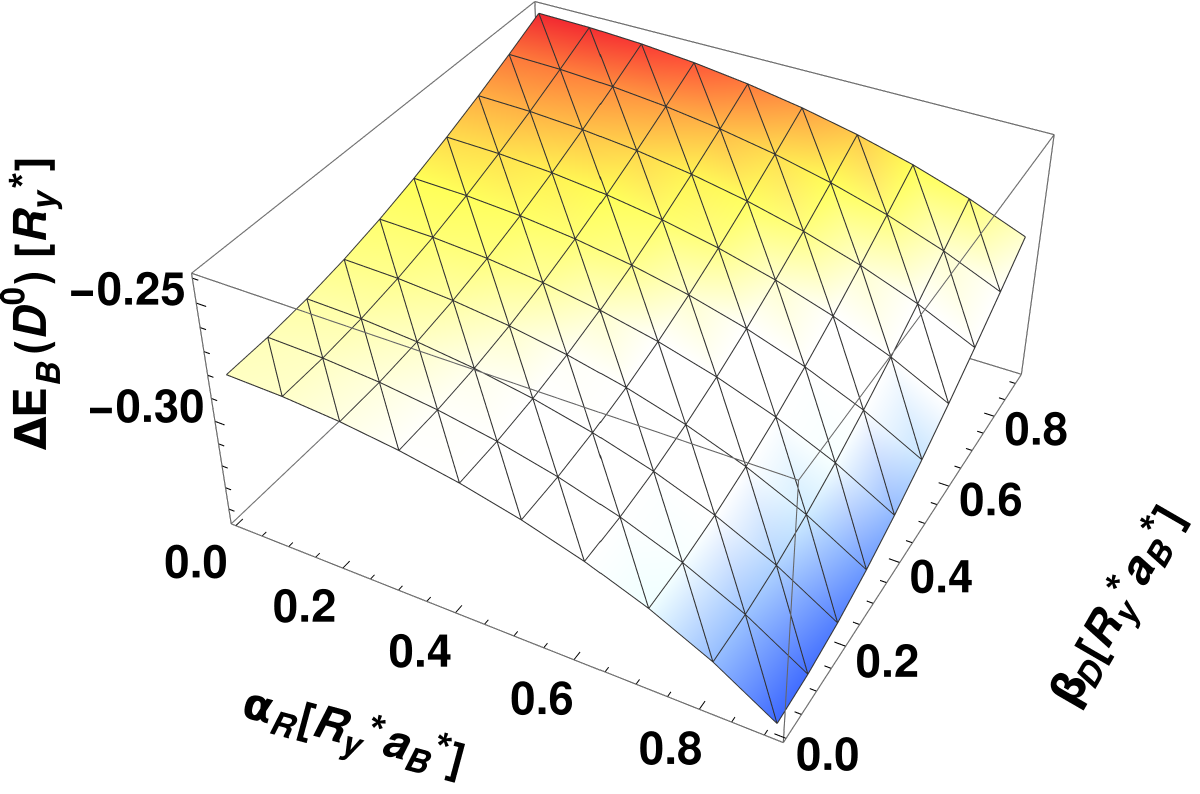}
	\caption{3D plot showing the combined influence of Rashba and Dresselhaus spin-orbit interactions on the relative binding energy$(1s\to2p)$ of an off-centered $D^0$ impurity confined in a Gaussian quantum dot.}
	\label{fig:del_eb_vs_ab}
\end{figure}
\\
\indent Fig-\ref{fig:del_eb_vs_ab} presents the variation of the relative binding energy between the $1s$ and $2p$ impurity states as a function of the Rashba and Dresselhaus SOI strengths. The negative sign of relative binding energy indicates that the impurity is weakly bound in the $1st$ excited state compared to the ground state which is expected. The relative binding energy becomes progressively more negative with increasing Rashba coupling, whereas it shifts towards less negative values as the Dresselhaus interaction is strengthened, demonstrating the competing influence of the two spin-orbit mechanisms. A similar dependence of donor binding energies on Rashba and Dresselhaus coupling strengths has also been documented in earlier investigations, covering both ground and excited-state properties\cite{Kachu2024}.
\begin{figure}[htbp]
	\centering
	\includegraphics[width=1.0\linewidth]{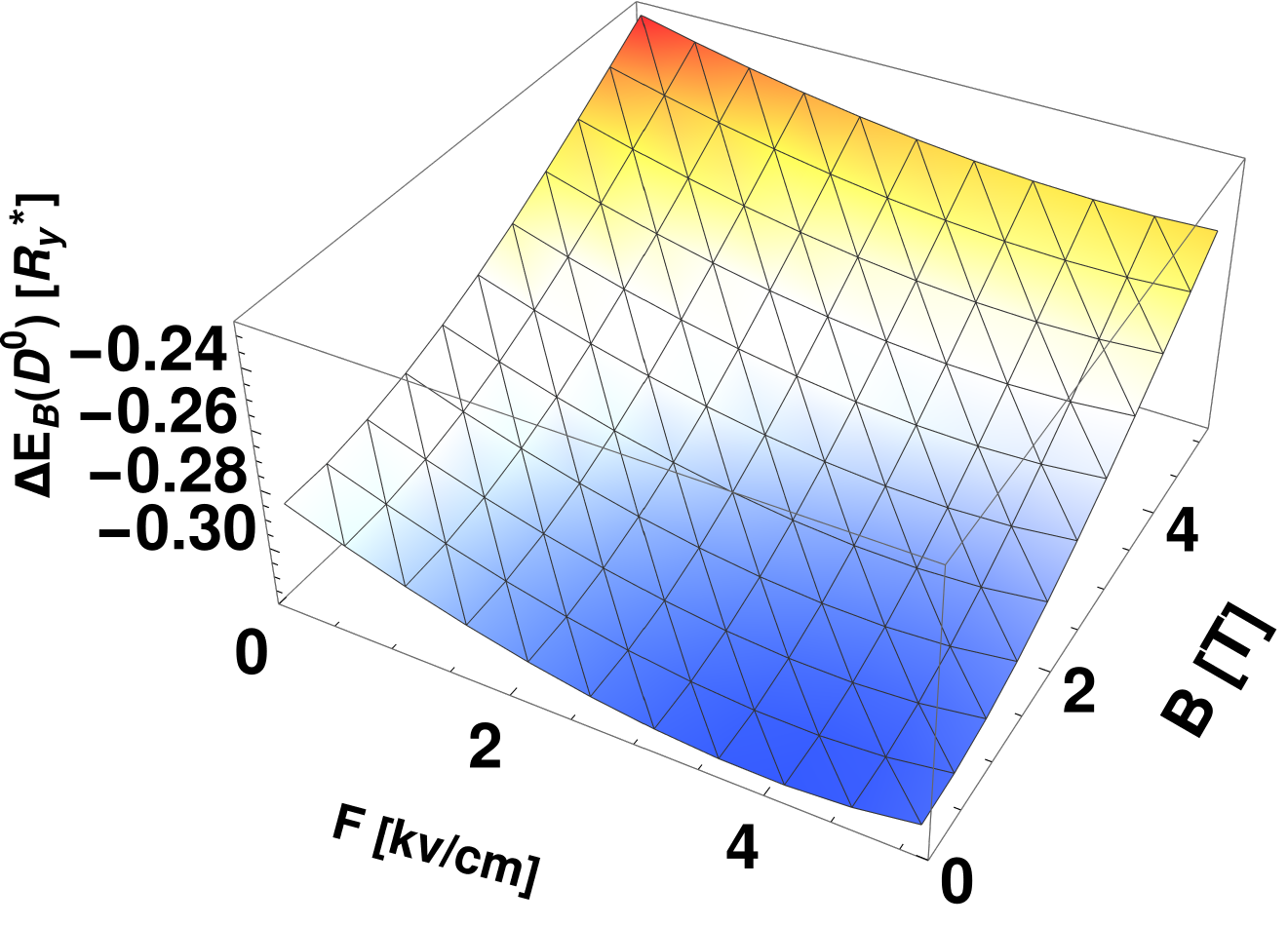}
	\caption{3D plot illustrating the combined influence of the electric field $(F)$ and magnetic field $(B)$ on the relative binding energy $(\Delta E_{B})$ of an off-centered $D^0$ impurity confined in a Gaussian quantum dot with $V_{0}=40, R=1.0a_{B}^{*}, D=0.8a_{B}^{*}, \alpha_{R}=0.5, \beta_{D}=0.5$.}
	\label{fig:del_eb_vs_EB}
\end{figure}
\\
\indent Fig-\ref{fig:del_eb_vs_EB} shows the variation of the relative binding energy under the simultaneous action of the electric$(F)$ and magnetic$(B)$ fields. An increase in the magnetic field shifts the relative binding energy towards less negative values, whereas the electric field produces the opposite effect by making it progressively more negative. This contrasting behaviour indicates that the two external fields modify the relative binding strengths of the $1s$ and $2p$ impurity states in different ways. The vally type nature at some intermediate electric and magnetic field strengths indicate that the difference of binding energies of the $1s$ and $2p$ state is minimum in that region. This behaviour arises from the competing influences of electric-field-induced polarization and magnetic confinement, which modify the spatial distributions of the ground and excited impurity states differently. The resulting field-dependent redistribution of the electronic wavefunctions, together with the re-optimization of the variational parameters, produces a non-linear variation of their relative binding characteristics. Thus, the curvature of the surface reflects the coupled response of the two impurity states to the external perturbations rather than an independent contribution from either field alone.
 A qualitatively consistent magnetic-field dependence was reported by Kachu \textit{et al.}, who examined the ground- and first excited-state binding energies of an off-centred donor in a Gaussian quantum dot without an external electric field. Their results indicated a more pronounced enhancement of the excited-state binding energy compared with its ground-state counterpart as the magnetic field increased. The increasing relative binding energy observed in our calculations under simultaneous electric and magnetic fields is consistent with this differential enhancement \cite{Kachu2024}.
\\
\begin{figure}[htbp]
	\centering
	
	\begin{minipage}[c]{0.48\linewidth}
		\centering
		\includegraphics[width=\linewidth,height=0.16\textheight]{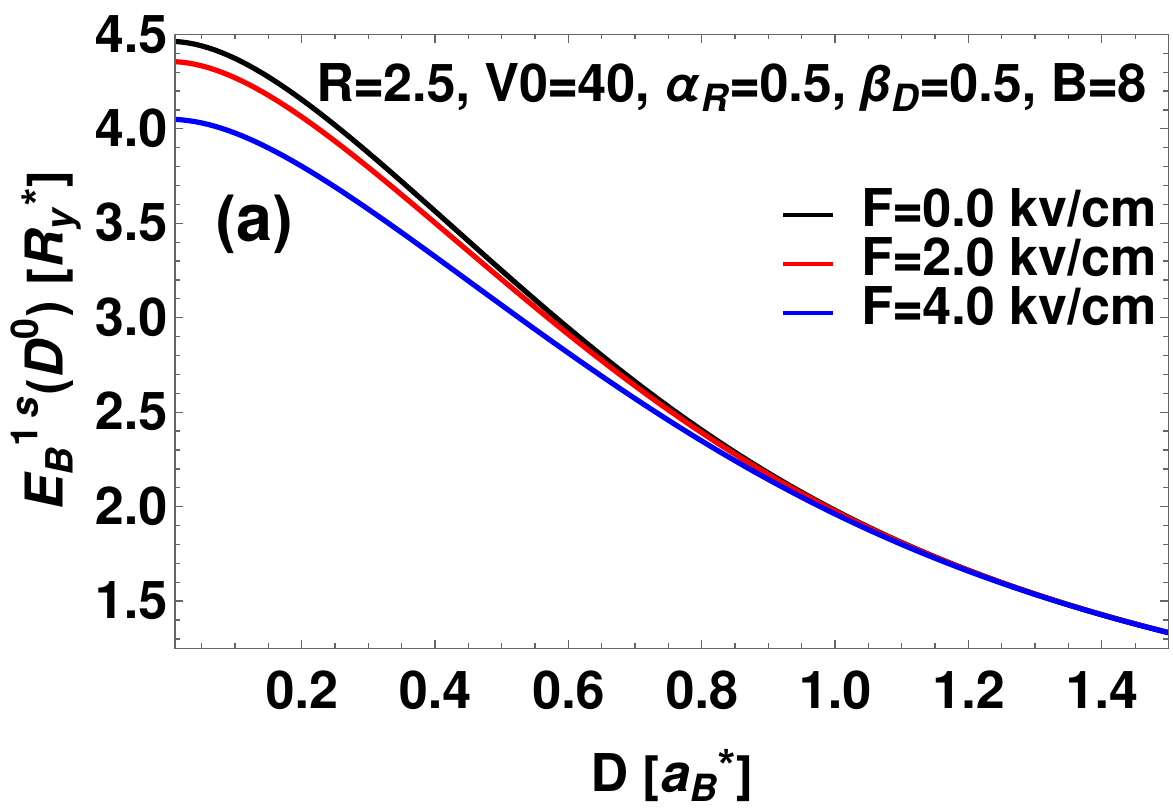}
	\end{minipage}
	\hfill
	\begin{minipage}[c]{0.48\linewidth}
		\centering
		\includegraphics[width=\linewidth,height=0.16\textheight]{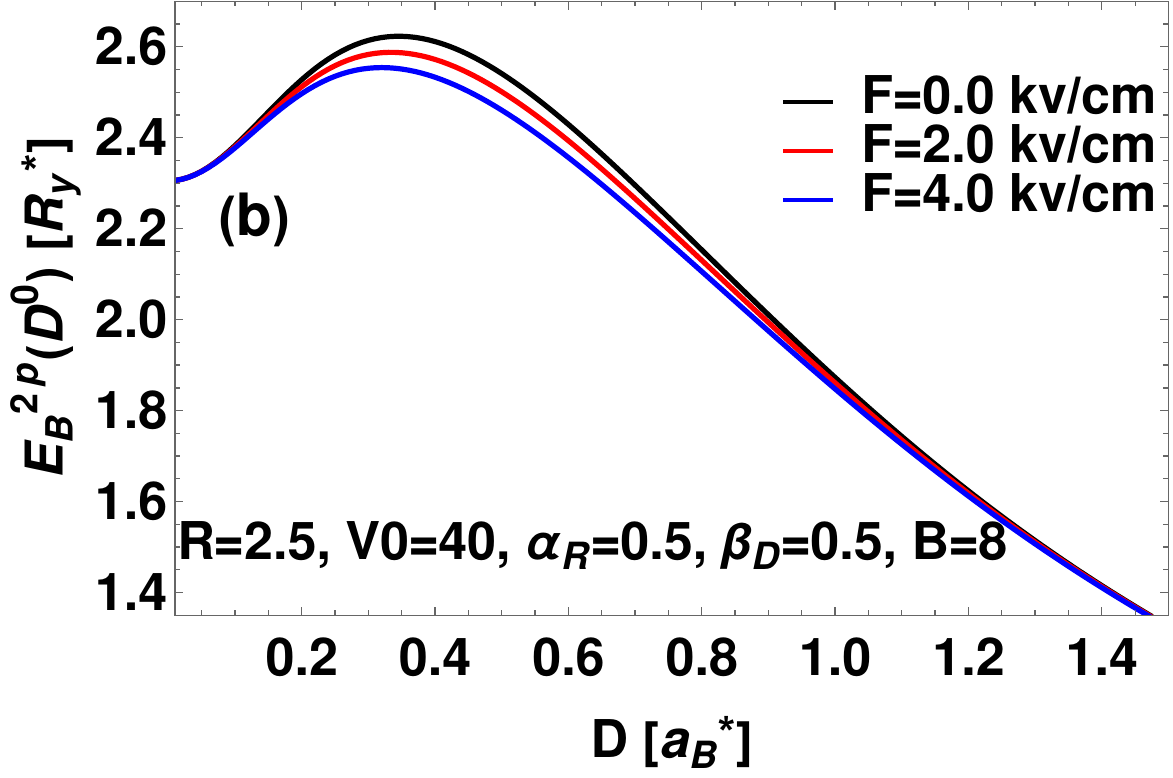}
	\end{minipage}
	
	\caption{Variation of the (a) ground-state $(1s)$ and (b) first excited-state $(2p)$ binding energies of an off-centered $D^0$ impurity with impurity displacement for different values of the electric field strengths$(F)$. }
	\label{fig:Eb_vs_D}
	
\end{figure}
\\
\indent Fig-\ref{fig:Eb_vs_D} depicts the variation of the ground $(1s)$[fig-\ref{fig:Eb_vs_D}.a] and first excited $(2p)$[fig:\ref{fig:Eb_vs_D}.b] state binding energies with impurity displacement for different electric field strengths$(F)$. The ground-state binding energy$(E_{B}^{1s})$ decreases monotonically as the impurity moves away from the dot center, reflecting the gradual weakening of the electron-impurity Coulomb attraction. The non-monotonic behaviour of the excited state$(2p)$ binding energy$(E_{B}^{2p})$ originates from the competition between the confinement potential and the electron-impurity Coulomb interaction. At small impurity displacements, the spatially extended nature of the 2p state allows the electron distribution to adjust to the displaced donor without a significant loss of confinement, resulting in a slight enhancement of the binding energy. As the impurity is moved further away from the dot center, the increasing electron-impurity separation weakens the Coulomb attraction, which eventually becomes the dominant effect and leads to a continuous reduction in the binding energy. The effect of the electric field is also suppressed at higher values of impurity displacement. 
 For comparison, the binding energies obtained in the present work could be compared with the results reported by Kachu et al. \cite{KACHU2024416376} in Fig. 3(consider $\gamma=2$ case) of their study, where the ground-state ($1s$) and first-excited-state ($2p$) binding energies of an off-center \(D^0\) impurity in a GaAs quantum dot were investigated as a function of the impurity position without any external electric field for different confinement shapes. The results of the present study exhibit consistently higher binding energies(For $F=0$ and $F\neq0$) for both the ground state and first-excited state compared with those reported in Ref. \cite{KACHU2024416376}. The higher binding energies obtained for both energy levels demonstrate the effectiveness of the present model in enhancing impurity localization relative to the corresponding results reported previously.
\\
\begin{figure}[htbp]
	\centering
	
	\begin{minipage}[c]{0.47\linewidth}
		\centering
		\includegraphics[width=\linewidth,height=0.14\textheight]{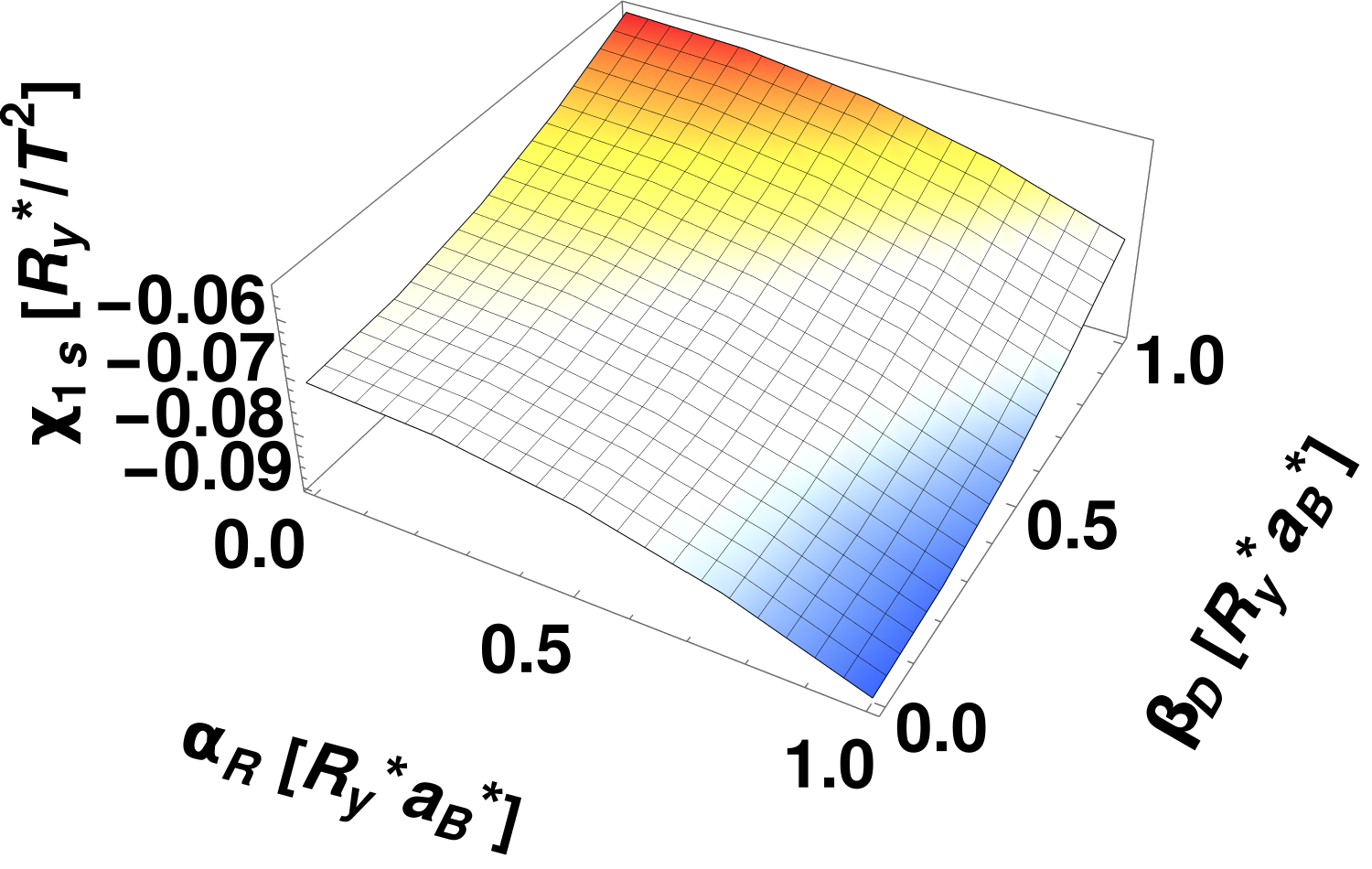}
		\vspace{2mm}
		{\small (a)}
	\end{minipage}
	\hfill
	\begin{minipage}[c]{0.47\linewidth}
		\centering
		\includegraphics[width=\linewidth,height=0.14\textheight]{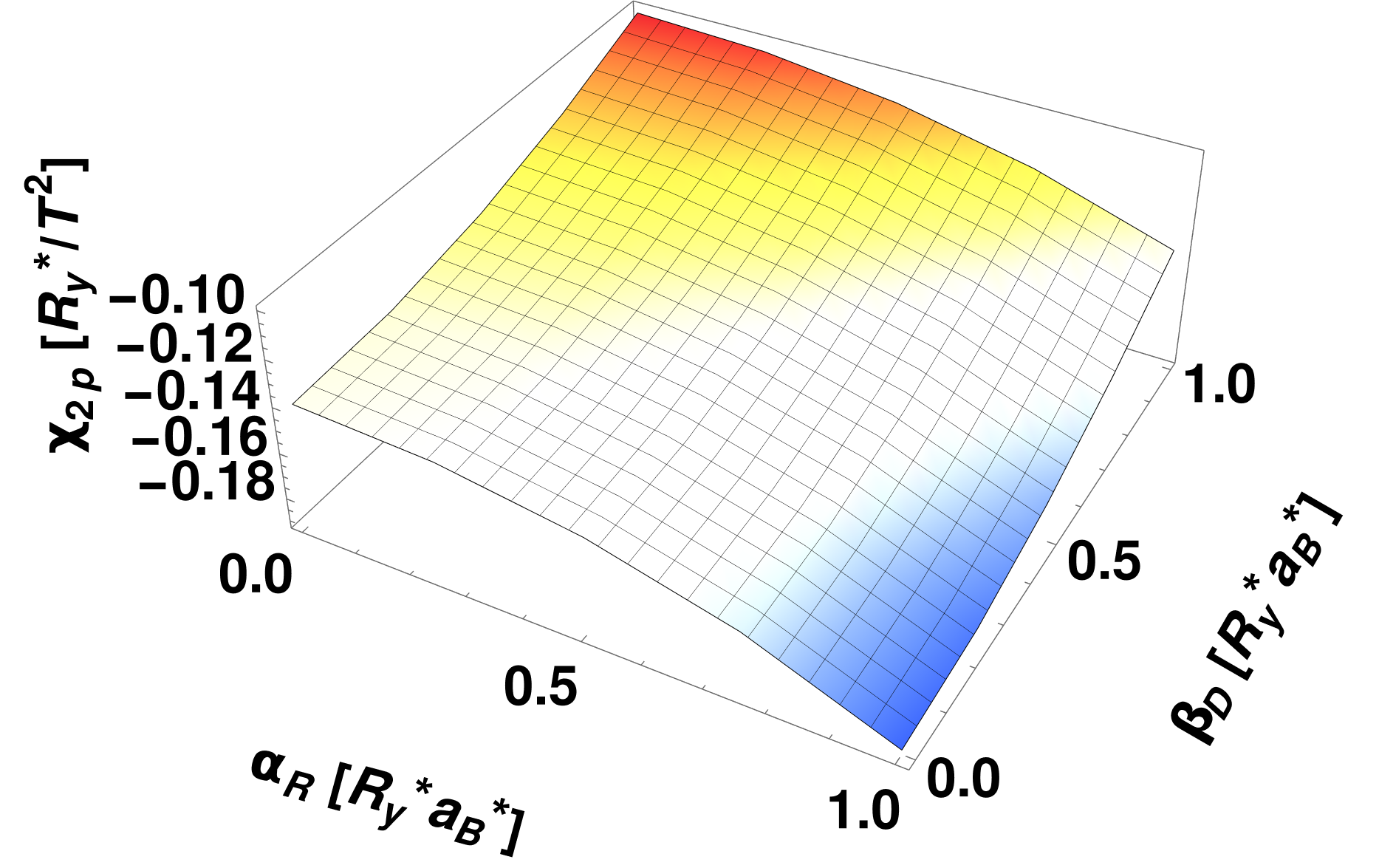}
		\vspace{2mm}
		{\small (b)}
	\end{minipage}
\caption{3D plots showing the dependence of the (a) ground-state  $(\chi_{1s})$ and (b) first excited-state $(\chi_{2p})$ magnetic susceptibilities of the off-centered $D^0$ impurity on the Rashba $(\alpha_{R})$ and Dresselhaus $(\beta_{D})$ spin-orbit coupling strengths with $V_{0}=40 R_{y}^{*}, B=5T, R=1.0 a_{B}^{*}, D=0.8 a_{B}^{*}, F=2.5 kv/cm$.}
\label{fig:S_vs_ab}
\end{figure}
\\
\indent Fig-\ref{fig:S_vs_ab} illustrates the dependence of the ground$(\chi_{1s})$[fig-\ref{fig:S_vs_ab}.a] and first excited-state$(\chi_{2p})$[fig-\ref{fig:S_vs_ab}.b] magnetic susceptibilities on the Rashba and Dresselhaus spin-orbit coupling strengths. In both states, the susceptibility remains negative throughout the investigated parameter range, confirming the diamagnetic character of the impurity system. Increasing the Rashba coupling drives the susceptibility towards more negative values, corresponding to an enhancement of the diamagnetic response, whereas a stronger Dresselhaus interaction produces the opposite trend by reducing its magnitude. The contrasting evolution of the magnetic susceptibility with Rashba and Dresselhaus coupling can be attributed to their different contributions to the field-dependent variational energy of the impurity complex. Since the susceptibility is determined by the second derivative of the energy with respect to the magnetic field, any spin-orbit-induced modification of the energy curvature directly affects the magnetic response. In the present off-centered configuration, the displacement of the donor breaks the spatial symmetry of the confinement and alters the electron-density distribution, thereby changing the relative influence of the two spin-orbit interactions. The Rashba contribution increases the magnitude of the negative energy curvature, resulting in a more pronounced diamagnetic response, whereas the Dresselhaus contribution reduces this curvature and consequently weakens the diamagnetism. The stronger variation observed for the ($2p$) state is consistent with its more extended and nodal spatial character, which makes the excited-state energy more responsive to spin-orbit-induced changes than the localized ($1s$) state.
\\
\begin{figure}[htbp]
	\centering
	
	\begin{minipage}[c]{0.47\linewidth}
		\centering
		\includegraphics[width=\linewidth,height=0.14\textheight]{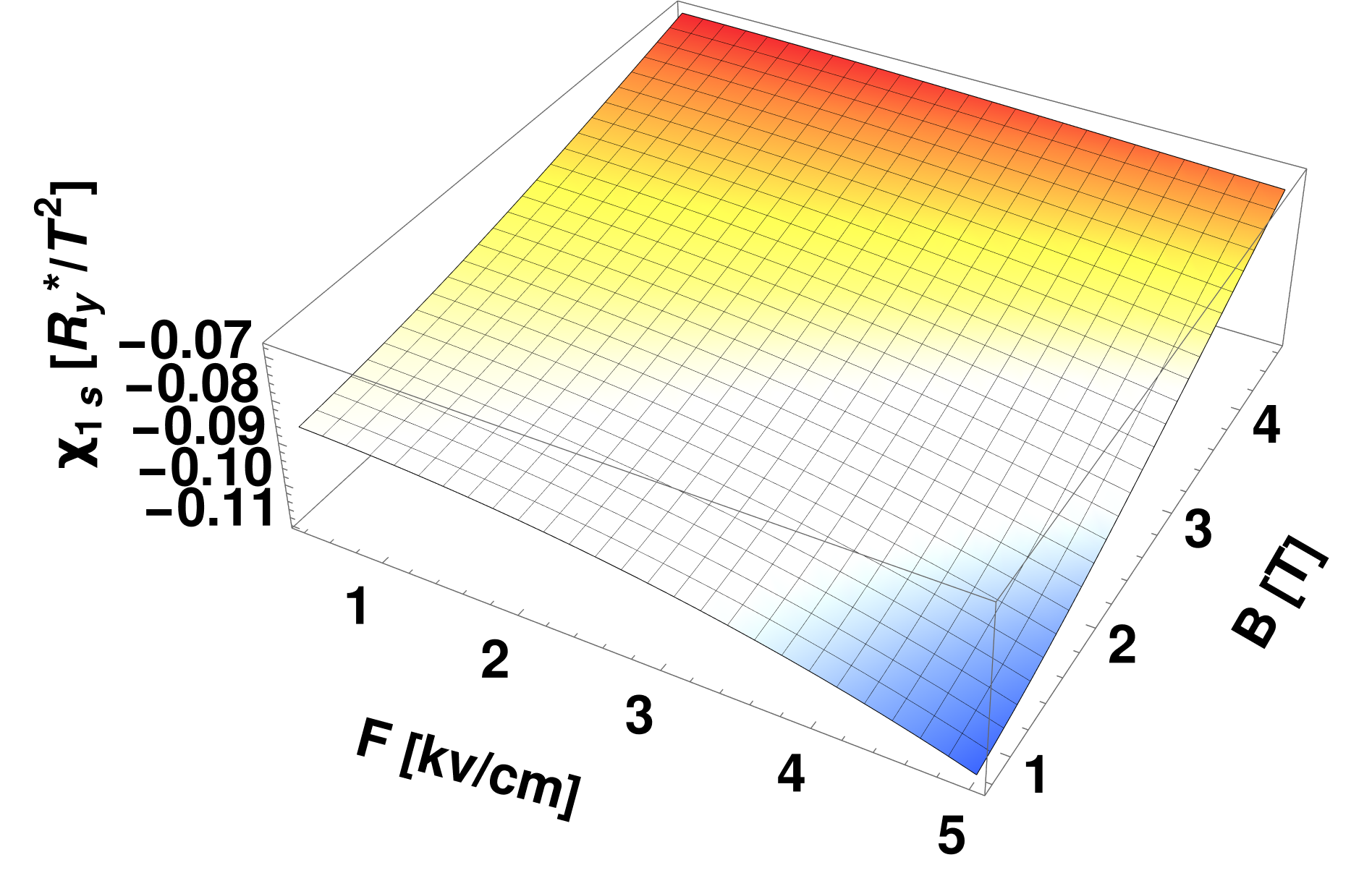}
		\vspace{2mm}
		{\small (a)}
	\end{minipage}
	\hfill
	\begin{minipage}[c]{0.47\linewidth}
		\centering
		\includegraphics[width=\linewidth,height=0.14\textheight]{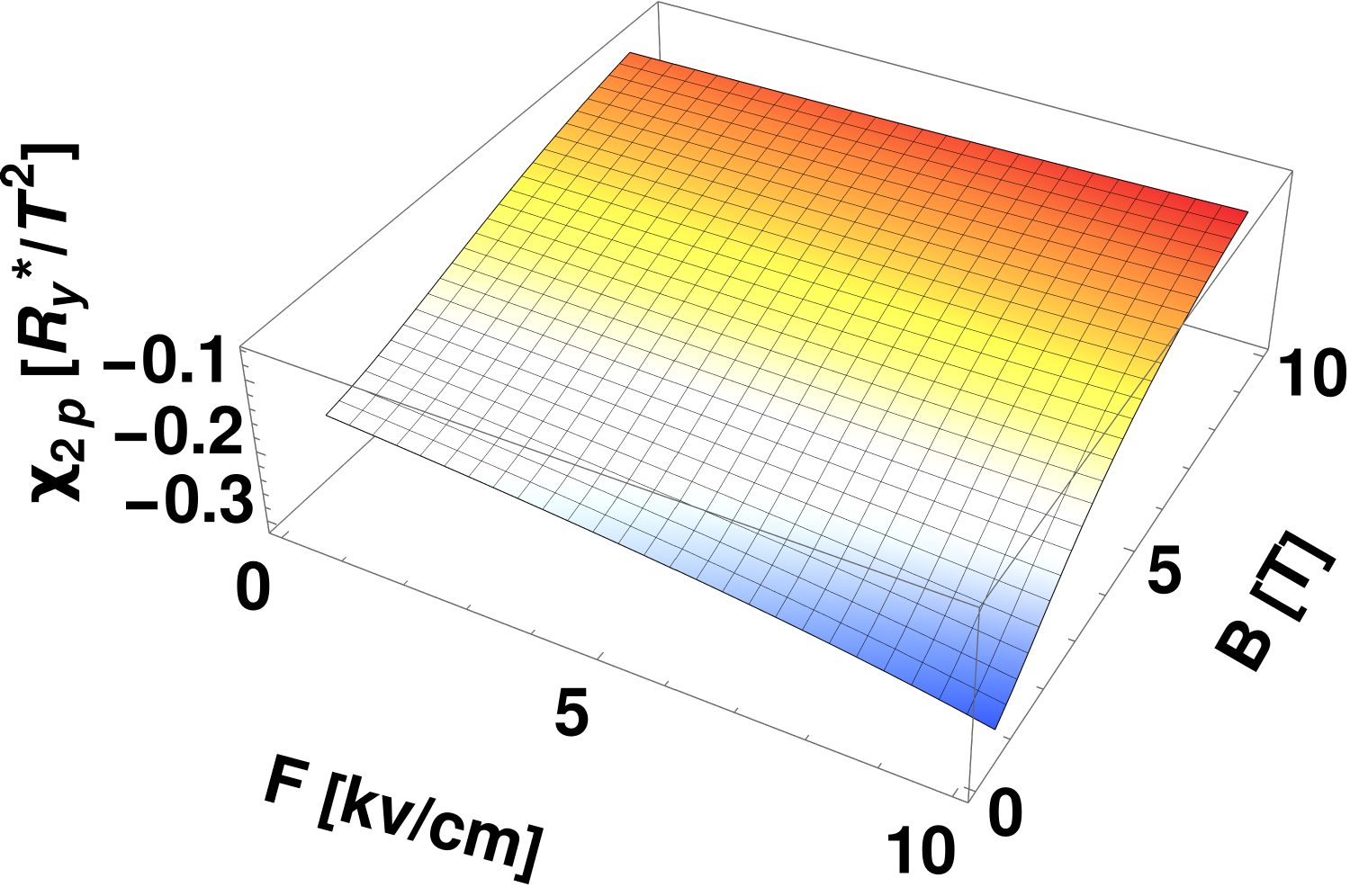}
		\vspace{2mm}
		{\small (b)}
	\end{minipage}
	\caption{Magnetic-field and electric-field dependence of the (a) ground-state \((\chi_{1s})\) and (b) first excited-state \((\chi_{2p})\) magnetic susceptibilities for an off-centred \(D^0\) impurity in a GaAs Gaussian quantum dot.}
	\label{fig:S_vs_EB}
\end{figure}
\\
\indent Fig-\ref{fig:S_vs_EB} illustrates the combined influence of the applied electric and magnetic fields on the magnetic susceptibilities of the ground $(1s)$[fig-\ref{fig:S_vs_EB}.a] and first excited $(2p)$[fig-\ref{fig:S_vs_EB}.b] states. The susceptibility remains negative throughout the investigated parameter range, confirming the diamagnetic nature of the impurity system. Increasing the magnetic field strength shifts the susceptibility towards less negative values, indicating a gradual weakening of the diamagnetic response. In contrast, the electric field enhances the magnitude of the susceptibility by driving it to more negative values. This opposing behaviour reflects the competition between magnetic confinement and electric-field-induced polarization of the impurity-bound electron. Moreover, the excited state exhibits similar variation over a more expanded region of electric and magnetic field compared to the ground state. Comparable magnetic-field-dependent behaviour of the ground-state susceptibility has also been documented in previous studies of centred donor impurities, both in the absence\cite{Saini2019} and presence\cite{BANERJEE2026174510} of an external electric field.
\\
\begin{figure}[htbp]
	\centering
	
	\begin{minipage}[c]{0.48\linewidth}
		\centering
		\includegraphics[width=\linewidth,height=0.16\textheight]{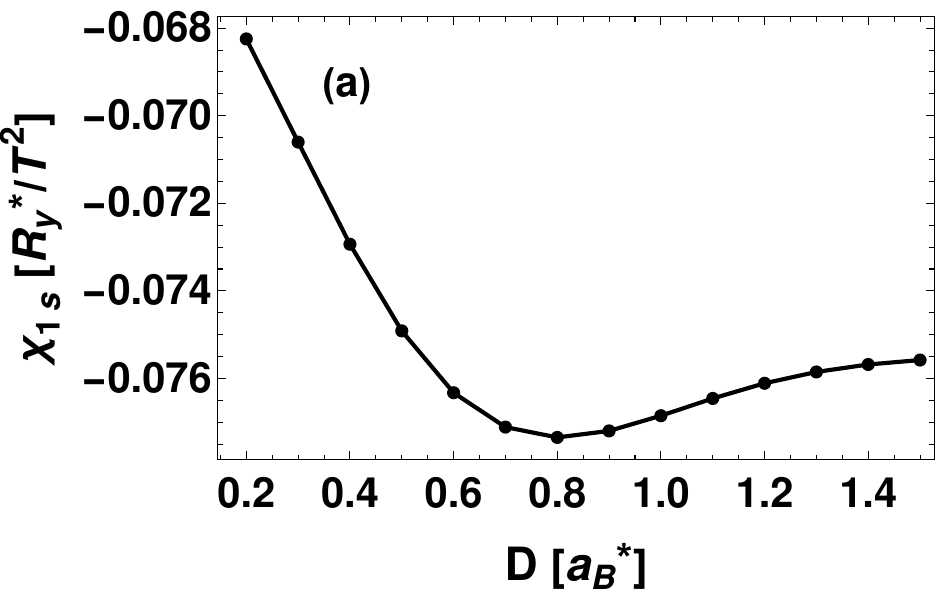}
	\end{minipage}
	\hfill
	\begin{minipage}[c]{0.48\linewidth}
		\centering
		\includegraphics[width=\linewidth,height=0.16\textheight]{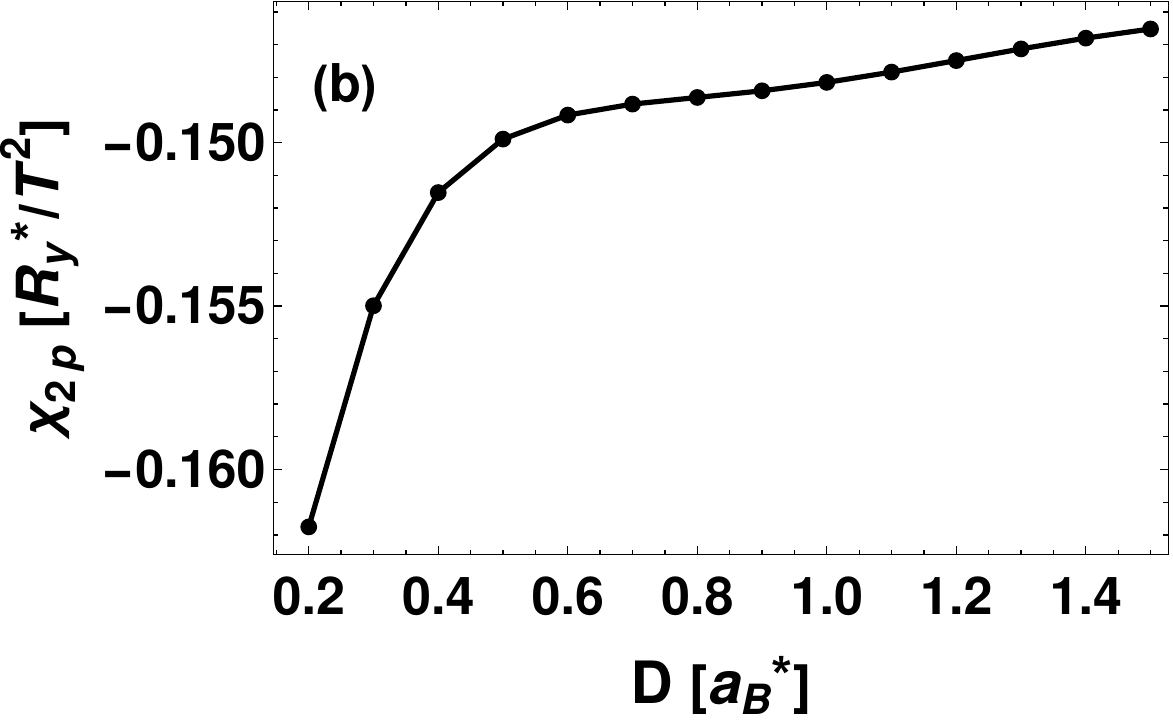}
	\end{minipage}
	
	\caption{Variation of the (a) ground-state and (b) first excited-state magnetic susceptibilities of an off-centered $D^0$ impurity with impurity displacement with $V_{0}=40 R_{y}^{*}, R=1.5 a_{B}^{*}, \alpha_{R}=0.5, \beta_{D}=0.5, F=2.5 kv/cm$. }
	\label{fig:S_vs_D}
	
\end{figure}
\\
\indent Fig-\ref{fig:S_vs_D} presents the variation of the magnetic susceptibilities of the ground $(1s)$[fig-\ref{fig:S_vs_D}.a] and first excited $(2p)$[fig-\ref{fig:S_vs_D}.b] states with impurity displacement. In both cases, the susceptibility remains negative, confirming the diamagnetic nature of the impurity system. The ground-state susceptibility, $\chi_{1s}$ initially becomes more negative with increasing displacement, reaches a minimum around $D\approx0.8a_{B}^{*}$, and then exhibits a slight recovery at larger displacements. The non-monotonic variation of the ground-state magnetic susceptibility with impurity displacement reflects the competing influences of donor localization, quantum confinement, and the magnetic response of the electronic distribution. At intermediate displacement, the redistribution of the electron probability density modifies its transverse spatial extent, leading to an enhanced diamagnetic response. Beyond this region, the competing effects of the displaced Coulomb potential and the dot-centred confinement alter the spatial distribution and produce a partial recovery of the susceptibility. Similar sensitivity of diamagnetic susceptibility to impurity location has been reported in strained type-I and type-II core/shell quantum dots, where the position of the donor determines the characteristic extremum in the magnetic response \cite{Perez2023}. Position-dependent modifications of magnetic susceptibility have also been demonstrated in semiconductor quantum rings, where changes in the impurity location influence the magnetic-field-dependent energy spectrum and associated magnetic response \cite{Shaer2023}. In contrast, the excited-state susceptibility, $\chi_{2p}$ increases monotonically towards less negative values, indicating a gradual weakening of the diamagnetic response. This distinct displacement dependence of the magnetic susceptibility in the two impurity states can be understood from their different spatial distributions and the resulting changes in their field-dependent energies. Because ($\chi=-\partial^2E/\partial B^2$), the displacement of the donor modifies the magnetic response through its effect on the balance between confinement, electron-impurity attraction, and the spatial redistribution of the bound electron. For the ($1s$) state, a small displacement initially enhances the magnitude of the diamagnetic response, leading to the minimum in ($\chi_{1s}$). At larger displacements, the weakening Coulomb attraction reduces this contribution and produces the observed recovery. The ($2p$) state, owing to its nodal and more spatially extended character, responds differently to the displaced impurity, resulting in a gradual shift of ($\chi_{2p}$) towards less negative values without an intermediate minimum. Thus, the contrasting susceptibility profiles reflect the state-dependent modification of the magnetic energy curvature induced by impurity off-centering.

\section{SUMMARY AND CONCLUSION}
In the present study, we have investigated an off-centred hydrogenic \(D^0\) donor in a GaAs Gaussian quantum dot using a variational effective-mass approach, incorporating simultaneous electric and magnetic fields alongside Rashba and Dresselhaus spin-orbit interactions. Particular emphasis was placed on the \(1s\) and \(2p\) states and their electronic and magnetic responses to external perturbations. Increasing impurity displacement raises the ground-state energy, whereas the excited state exhibits a non-monotonic trend. This contrasting behaviour originates from their distinct spatial structures and unequal responses to the displaced Coulomb potential, with the nodal and extended character of the \(2p\) state, accounting for its greater sensitivity to impurity position.

This state-dependent response is also reflected in the binding and transition energies. With increasing impurity displacement, the \(1s\) binding energy decreases monotonically, whereas the \(2p\) binding exhibits a non-monotonic variation arising from the competition between confinement, Coulomb attraction, and wavefunction redistribution. The \(1s\rightarrow2p\) transition energy increases with magnetic field but generally decreases with electric field, while the relative binding energy remains sensitive to their combined influence. These observations demonstrate the extent to which impurity position, confinement, and field-induced polarization govern the energetic characteristics of the donor complex. 
In a broader context, Fainblat \textit{et al.} highlighted the importance of carrier-impurity exchange coupling in determining the spin-dependent and magneto-optical properties of magnetic-ion-doped colloidal quantum dots \cite{Fainblat2017}.
Alongside these field-dependent energetic changes, Rashba and Dresselhaus interactions introduce distinct modifications to the donor states. Increasing Rashba coupling lowers the energies of both states($1s$ and $2p$), whereas Dresselhaus coupling produces a weaker correction with the opposite tendency for the off-centred configuration. This contrast is more pronounced for the \(2p\) state owing to its characteristic angular and nodal structure. 
Comparable SOI-induced changes in impurity spectra and spin degeneracy have been reported by Esmaeili \textit{et al.} for core-shell quantum dots \cite{Esmaeili2018}, while Chattopadhyay \textit{et al.} demonstrated the role of exchange and Zeeman interactions in spin-selective transport \cite{Chattopadhyay2014}. Regarding the magnetic response, the susceptibility remains negative throughout the investigated range, confirming diamagnetic behaviour. Its magnitude, however, depends noticeably on impurity displacement, external fields, and SOI strengths. Notably, the \(1s\) susceptibility varies non-monotonically with donor position, whereas the \(2p\) response evolves more systematically towards weaker diamagnetism as the impurity moves away from the dot centre.
\\ \indent Taken together, these findings establish impurity displacement as an effective means of modifying the excited-state characteristics and magnetic response of a donor-bound quantum dot. Its combined use with electric and magnetic fields and Rashba-Dresselhaus interactions offers multiple avenues for controlling the energy spectrum, binding characteristics, transition energies, and magnetic properties. The pronounced sensitivity of the \(2p\) state to these parameters may provide useful opportunities for developing tunable semiconductor nanostructures and spin-dependent quantum devices.
\\
The present investigation is limited to a single donor impurity within the effective-mass and variational framework, with the electronic properties represented by selected trial wavefunctions for the \(1s\) and \(2p\) states. Effects arising from atomistic interface structures, conduction-band non-parabolicity, and finite temperature have not been considered. Incorporating these aspects, together with a more comprehensive treatment of higher excited states, would provide interesting directions for future investigations.
\section{Appendix}
\setcounter{figure}{0}
\renewcommand{\thefigure}{A\arabic{figure}}
\begin{figure}[H]
	\centering
	\begin{minipage}[c]{0.48\linewidth}
		\centering
		\includegraphics[width=\linewidth,height=0.15\textheight]{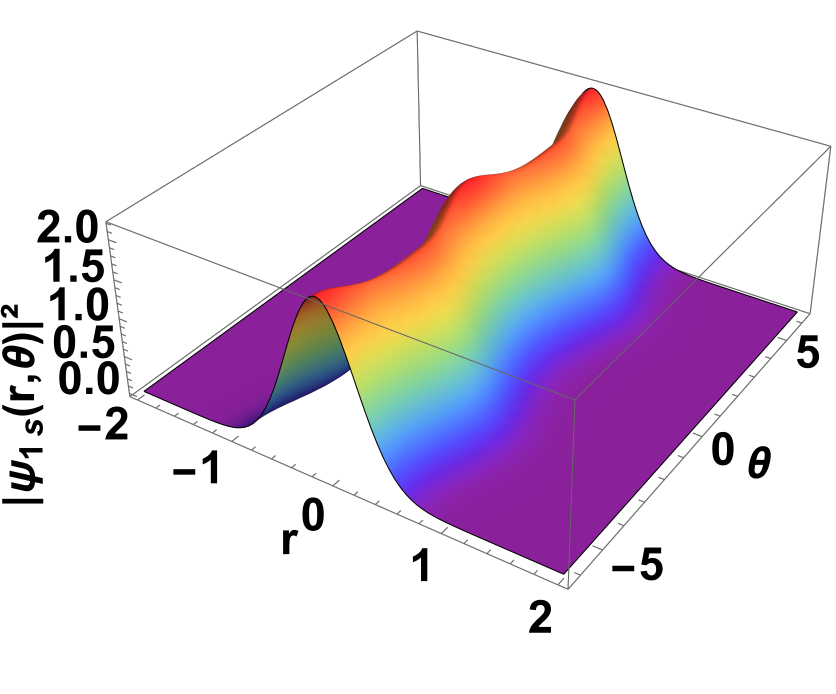}
		{\small (a)}
	\end{minipage}
	\hfill
	\begin{minipage}[c]{0.48\linewidth}
		\centering
		\includegraphics[width=\linewidth,height=0.15\textheight]{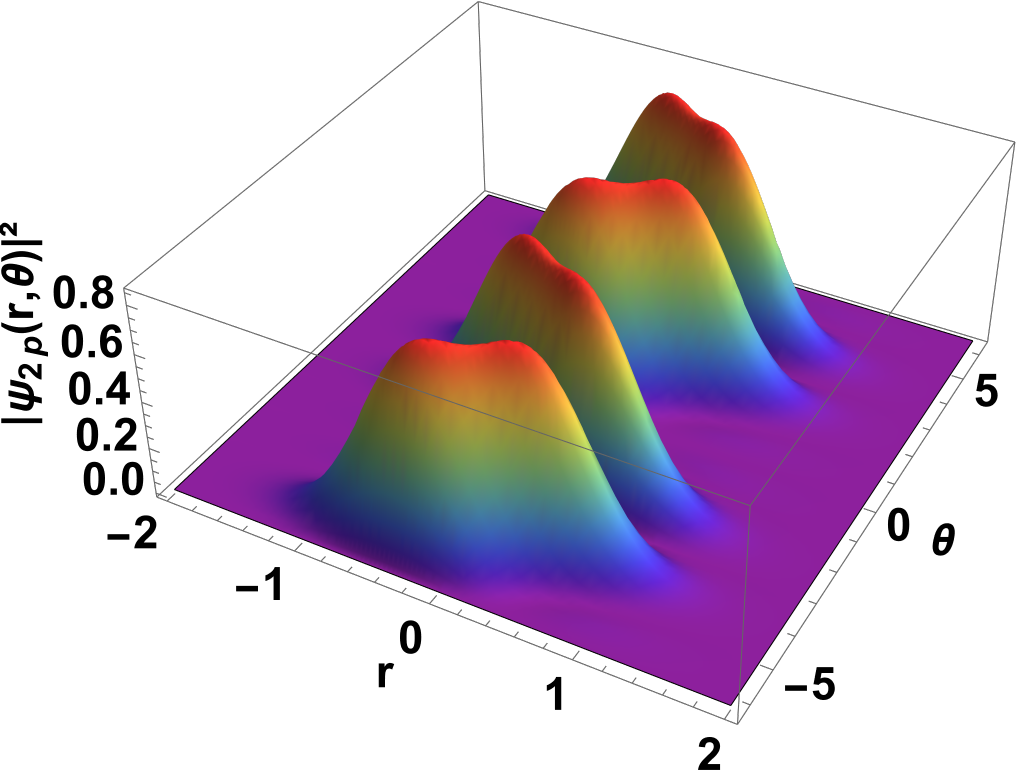}
		{\small (b)}
	\end{minipage}
	
	\caption{Three-dimensional probability-density distributions of the (a) ground-state (\(1s\)) and (b) first excited-state (\(2p\)) wave functions of the off-centered \(D^0\) impurity in a Gaussian quantum dot with $V_{0}=40\ R_{y}^{*}, R=1.5\ a_{B}^{*}, \alpha_{R}=0.5, \beta_{D}=0.5, B=10\ T, F=2.5\ kv/cm$. }
	\label{fig:Prob_density}
	
\end{figure}
The spatial probability densities of the ground (\(1s\)) and first excited (\(2p\)) impurity states are shown in Fig.\ref{fig:Prob_density}, respectively, for the representative parameters used in the present calculation. The \(1s\) state exhibits a strongly localized and predominantly single-peaked distribution, indicating that the electron remains concentrated within a restricted region of the quantum dot. In contrast, the \(2p\) state possesses a distinctly broader and multi-lobed probability distribution accompanied by regions of strongly reduced density, reflecting its nodal and angular character. The pronounced difference between the two distributions demonstrates that the excited state is spatially more extended and structurally more sensitive to the symmetry-breaking effects associated with the off-centered impurity. Consequently, changes in impurity position or external fields can modify the \(2p\) probability distribution more substantially than the ground state, providing a spatial interpretation of the different displacement and field dependences observed for the two energy levels.
\section*{Acknowledgements}
Pooja Saini gratefully acknowledges the financial support from the ANRF-NPDF fellowship under Grant No. (ANRF/PAIR/2025/000021/PAIR-A).

\end{document}